\documentclass[nolinenumbers]{aastex631}

\usepackage{newtxtext,newtxmath}

\usepackage[T1]{fontenc}

\usepackage{amsmath}
\usepackage{graphicx}	
\usepackage{svg}    
\usepackage{makecell}   %
\usepackage{cuted}
\usepackage{amssymb}	
\usepackage{hyperref}
\usepackage{threeparttable}
\usepackage{booktabs}
\usepackage{float}
\usepackage{etoolbox}

\begin{document}

\title[Potential to Detect Nearby 'Earth']{The Potential of Detecting Nearby Terrestrial Planets in the HZ with Different Methods}

\author{Hao Qiao-Yang}
\affiliation{School of Astronomy and Space Science, Key Laboratory of Ministry of Education, \\
Nanjing University, Nanjing, 210023, China}

\author{Zhou Shen-Wei}
\affiliation{School of Astronomy and Space Science, Key Laboratory of Ministry of Education, \\
Nanjing University, Nanjing, 210023, China}

\author{Liu Hui-Gen}
\thanks{E-mail: huigen@nju.edu.cn}
\affiliation{School of Astronomy and Space Science, Key Laboratory of Ministry of Education, \\
Nanjing University, Nanjing, 210023, China}



\begin{abstract}
Terrestrial planets in the habitable zone (hereafter HZ) around nearby stars are of great interest and provide a good sample for further characteristics of their habitability. In this paper, we collect a nearby star catalog (NSC) within 20 pc according to the Gaia Catalog of Nearby Stars, (\citealt{2021AA...649A...6G}), complete the physical parameters of the stars, and select stars that are not brown dwarfs or white dwarfs. After selection, a sample of 2234 main-sequence stars is used to estimate the extended HZ. Then we inject Earth-like planets into the extended HZ around each star and calculate the signals with four methods, i.e. velocity amplitude for radial velocity, transit probability and depth for transit, stellar displacements for astrometry, and contrast and angular separation for imaging. Considering a typical noise model based on classic instruments, e.g. ESPRESSO, Kepler, Gaia, HabEx, and LIFE, we predict the highest possible detection number of Earth-like planets via different methods in the best-case hypothetical scenario. According to this, we conclude that both astrometry and imaging have the potential to detect nearby Earth-like planets around G-type stars, while radial velocity has the potential to detect 2\% of nearby Earth-like planets around M stars under a precision of $0.2 m/s$. Our work also provides the precision requirements for future missions to reveal the nearby Earth-like planet in the HZ.

\end{abstract}



\section{Introduction}

The exoplanet population has been growing fast since \citet{1995Peg51b} detected Peg 51 b via radial velocity (RV). More than 5400 exoplanets have been detected (up to 2023 Jun 14), most of which are contributed via transit methods by Kepler and TESS. However, few terrestrial planets in the habitablr zone (HZ) are found due to observational bias. According to the Catalog of Habitable Zone Exoplanets (\citealt{2023AJ....165...34H}),
 16 Earth-like planets with radii of 0.5-1.5 $R_{\oplus}$ are detected, whose orbits are always inside the optimistic HZ (hereafter Earth-like planets). There are eight temperate Earth analogs around nearby stars within 20 pc, e.g., TRAPPIST-1 d \citep{TRapPIST-12017}. The closest, Proxima b \citep{ Anglada_Escud__2016, Proxima2021}, is not included because its eccentricity is set as 0.35; therefore, it is not always located in the HZ. According to the NASA Exoplanet Archive (\href{https://exoplanetarchive.ipac.caltech.edu/cgi-bin/TblView/nph-tblView?app=ExoTbls&config=PS}{NASA Exoplanet Archive: Planetary Systems\footnote{\href{https://exoplanetarchive.ipac.caltech.edu/cgi-bin/TblView/nph-tblView?app=ExoTbls&config=PS}{https://exoplanetarchive.ipac.caltech.edu/cgi-bin/TblView/nph-tblView?app=ExoTbls\&config=PS}}}), ~85.9\% of terrestrial planets with radii of 0.5-1.5 $R_{\oplus}$ are over 100 pc from the Earth, and ~99.8\% of them are detected via transit.


Searching for planets around nearby stars is of great interest. Firstly, nearby brighter stars are more suitable for high-precision detecting; e.g. transit can achieve better precision on bright stars. Bright hosts are also suitable for RV follow-up. In contrast, Kepler stars, most of which are fainter than $Kp=14$, are not easy to follow up with the RV method. Nearby solar-like stars within 10 pc are usually brighter than $\rm V_{mag}\sim$ 5. Thus, the planetary mass around these stars can be constrained by RV data. Furthermore, to characterize the planetary atmosphere, we also need a high-precision transmission spectrum, which prefers bright nearby stars, to constrain more details about the climates and chemistry of planets (e.g. \citealt{2022Natur.609..229H}, \citealt{2023arXiv230104191L}). Thus detecting planets around nearby stars becomes more and more important for studying the habitability of exoplanets. 

The TESS mission, which focuses on transiting planets around nearby bright stars \citep{2014JAVSO..42..234R}, had detected 350 exoplanets and 6599 candidates as of 2023 June 14, with 310 out of 350 confirmed planet hosts brighter than $\rm T_{mag}=12$. TESS also detected several Earth-like planets, e.g. TOI-1266 c \citep{2020AJ....160..259S} and TOI 700 e \citep{2023arXiv230103617G}.

In this paper, we focus on nearby stars within 20 pc. Solar-like stars at 20 pc are bright enough ($\rm V_{mag}\sim6.5$), and are possible to detect terrestrial planets. Furthermore, the distance of 20 pc also corresponds to a moderate and achievable spatial separation for Sun-Earth analogous systems, with an angular separation of $\sim0.^{''}05$, which is approximate to the spatial resolution for JWST at 1 $\mu m$, although JWST can hardly directly image the planet due to the high contrast. \citet{2021AA...649A...6G} provided the Gaia Catalogue of Nearby Stars (GCNS), which includes stars within 100 pc, with high-reliability distances via Gaia EDR3. According to the GCNS, there are 2575 stars (including white dwarfs, WDs and brown dwarfs) within 20 pc. The exoplanet occurrence rates vary with the hosts' properties; according to \citep{2021AJ....161...36B}, the occurrence rate of planets in the HZ with radii between 0.5 and 1.5 $R_{\oplus}$, orbiting stars with effective temperatures between 4800 and 6300K is between $0.37_{-0.21}^{+0.48}$ and $0.60_{-0.36}^{0.90}$, in 68\% credible intervals. For M dwarfs, the cumulative planet occurrence rate is $2.5\pm0.2$ planets per star with planetary radii of 1-4 $R_{\oplus}$ and periods shorter than 200 days, estimated by \citep{2015ApJ...807...45D}. Thus, within a distance of 20 pc, there should be more than hundreds of habitable planets. However, only 16 Earth-like planets in the HZ are confirmed (\citealt{2023AJ....165...34H}), which is probably due to the strict geometry configuration requirements in the transit method and extreme precision requirements in the RV method.

We try to simulate the planetary signals based on the GCNS to compare with the typical observational precision of four typical methods: transit, RV, astrometry, and direct imaging. We want to investigate the potential of different methods to detect nearby Earth-like planets. The results will also benefit us in knowing the typical precision requirements for further observation to detect Earth-like planets and to optimize an observational strategy for further exoplanet missions.

 This paper is organized as follows. In section \ref{sec:complete}, we obtain the NSC within 20 pc of the GCNS and improve the parametric completeness of the catalog through cross-matching and empirical estimation. In section \ref{sec:HZsetting}, we introduce how we consider the HZ around the selected nearby stars and set the Earth-like planet in our simulations. The calculated signals due to four different methods are presented in section \ref{sec:Signal}. Then we statistically compare the potential of detecting nearby Earth-like planets according to the precision models in different methods in section \ref{sec:Detect}. Finally, the results are summarized and discussed in section \ref{sec:sum}.

\section{Complete parameters of nearby stars}\label{sec:complete}

This section introduces how we obtain a more parametric-complete catalog of nearby stars within 20 pc. Then, we also introduce how we obtain the stellar parameters by cross-matching or empirical correlations, which are used to estimate the HZ. In the end, we show the statistical results of the selected nearby stars.

\subsection{Collection of Basic Physical Parameters from other catalogs}\label{sec:BPPcollect}

To calculate the HZs of different stars, we need to know the physical parameters of the stars, e.g. the luminosity, temperature, radius, etc. Some crucial parameters are missing for a large fraction of stars in the GCNS. In this section, we are trying to complete the basic physical parameters of nearby stars, including mass, surface gravity, effective temperature, and radius. Meanwhile, other parameters, like metallicity, etc., are also collected when possible. From these physical parameters, we can demonstrate the stars in an H-R diagram and inspect the evolution stages of the stars. Since we only select stars via parallax, stellar parameter collection helps us select the main-sequence stars rather than the evolved ones. Two methods are adopted to complete the stellar parameters.

When choosing the catalogs used to complete the parameters, we use the Unified Column Descriptor on the Vizier website (UCD) to search for catalogs containing the five parameters: radius, surface gravity, mass, metallicity, and effective temperature. During UCD searching, we can acquire 4901 catalogs with rows of radius/surface gravity/mass/metallicity/effective temperature. It is not feasible for us to cross-match all of these catalogs with the GCNS. So, we have to choose a moderate number. Here we choose the top 100 catalogs via popularity, which are more frequently cited and relatively more popular or convincing. Additionally, we also select catalogs containing $\ge 500$ sources and having information on position and distance (or parallax), which can be used to cross-match with the GCNS. Eventually, 23 catalogs are 
reserved to complement stellar parameters.

Of the 23 catalogs, four are old versions and can be replaced by other catalogs like Gaia DR2 or DR3; six do not overlap with the GCNS, and two are catalogs of galaxies. Therefore, we exclude these 12 catalogs. Besides, we searched by keyword "nearby" on NASA ADS and added J/AJ/158/56 (2019) to our complement list. We also add LAMOST DR8 to collect stellar parameters. Finally, the parameter collection of NSC contains 14 catalogs (see Table~\ref{tab:catalog list}). Note that I/345/gaia2 and I/355/gaiadr3 are denoted as Gaia, and IV/38/tic is denoted as TIC.

\begin{minipage}[c]{\linewidth}
\centering
\begin{threeparttable}
    \caption{List of 14 catalogs Used to Complement the Basic Physical Parameters}
    	\label{tab:catalog list}
    	   \begin{tabular}{ccc} 
    		\hline
    		Catalog Name & Reference & Main parameter\\
    		\hline
    		II/224/cadars & \cite{2001AA...367..521P} & \emph R\\
    		J/apJS/168/297/table2 & \cite{2007ApJS..168..297T} & \emph M\\
    		V/130/gcs3 & \cite{2009AA...501..941H} & \emph M[H]\\
    		V/137D/XHIP & \cite{2012AstL...38..331A} & \emph M[H]\\
    		III/279/rave\_dr5 & \cite{2017AJ....153...75K} & \emph M[H]\\
    		I/345/gaia2 & \cite{2018AA...616A...1G} & $\rm T_{eff}$\\
    		VI/156/mdwarfp & \cite{2019MNRAS.489.2615M} & \emph R\\
    		J/AJ/158/56 & \cite{2019AJ....158...56H} & \emph M[H]\\
    		I/349/starhorse & \cite{2019AA...628A..94A} & \emph M\\
    		B/pastel/pastel & \cite{2016AA...591A.118S} & \emph M[H]\\
    		III/284/allstars & \cite{2020AJ....160..120J} & \emph M[H]\\
    		IV/38/tic & \cite{2019AJ....158..138S} & \emph R\\
    		LAMOST DR8 & LAMOST Collaboration\footnote{01;\href{http://www.lamost.org/dr8/v1.0/}{http://www.lamost.org/dr8/v1.0/}} & \emph M[H],log(\emph g),$T_{\rm eff}$\\
    		Gaia DR3 & \cite{2022arXiv220800211G} & \emph M[H],log(\emph g),$T_{\rm eff}$\\
    		\hline
	       \end{tabular}
        \begin{tablenotes}
            \item \textbf{Notes.} "Main Parameter" refers to the parameter mostly completed by a certain catalog.
        \end{tablenotes}
\end{threeparttable}
\end{minipage}

We can match the GCNS with the catalogs in Table \ref{tab:catalog list} using the CDS cross-match on Vizier. We uniformly use a cone search with a positional radius of $5{''}$. Then we combine the cross-match result with the GCNS according to the serial number of each star in the GCNS to complete the missing parameters. If a star is matched with multiple objects in different catalogs, we adopt the parameters in the most recent catalogs. In some cases, stars may have multiple matching sources due to the positional criterion of $5{''}$, and we adopt the parameters of the best-matched source. These stars make up less than 10\% of the whole catalog.

\begin{figure}
\centering
\includegraphics[width=0.9\textwidth,height=0.45\textwidth]{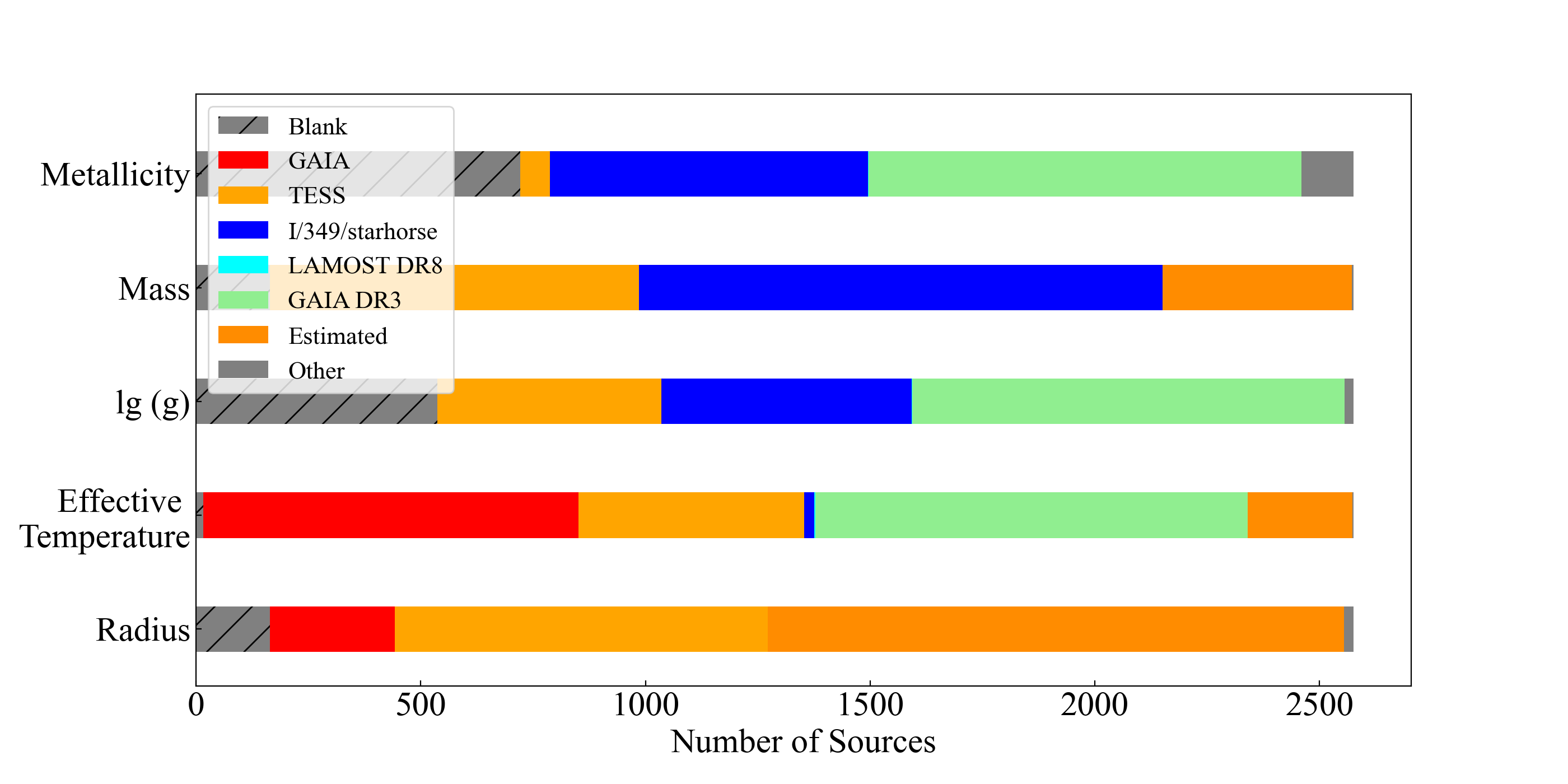}
\caption{The basic physical parameters in the NSC are from 14 catalogs, among which GAIA, TESS, and I/349/starhorse contribute the most. Other catalogs are symbolized by gray, and estimated parameters are shown in dark orange.}
\label{fig:source}
\end{figure}

The catalog I/349/starhorse (hereafter starhorse) derived from Gaia by \cite{2019AA...628A..94A} has a major contribution of the metallicity, mass, and surface gravity. This catalog was released in 2019 and successfully combines data from Gaia DR2, Pan-STARRS1, 2MASS, and AllWISE, optimizing the extinctions and astrophysical parameters for Gaia DR2 stars brighter than \emph{G} = 18 \citep{2019AA...628A..94A}. This catalog contributes 0.9\%, 22.8\%, 47.7\%, and 29.0\% of the information on stellar effective temperature, surface gravity, mass, and metallicity, respectively. LAMOST DR8 also contributes to the effective temperature, mass, and metallicity of two stars; see Figure \ref{fig:source}.

\subsection{Derivation of Basic Physical Parameters from empirical correlations}\label{sec:BPPempirical}

Since the basic parameters of some stars are absent, we derived a method to complete them with empirical formulae. As the original GCNS contains WDs, we will also exclude them while estimating the parameters. Note that we have not excluded brown dwarfs in this step, which will be done afterward. We use \emph{L}, \emph{M}, \emph{R}, $T_{\rm eff}$ to represent the luminosity, mass, radius, and effective temperature, respectively. When estimating the parameters, the following empirical formulae are used, including the luminosity-mass relation and radius-mass relations \citep{1991ap&SS.181..313D}:

\begin{equation}
    L \cong 1.18(M/M_{\odot})^{3.70}, 0.1  \leq M/M_{\odot} \leq 18.1
    \label{eq:LMR}
\end{equation}

\begin{equation}
    R \cong 1.01 (M/M_{\odot})^{0.724}, 0.1 \leq (M/M_{\odot}) \leq 18.1
    \label{eq:RMR}
\end{equation}

\begin{center}
\begin{threeparttable}
    \caption{When Completing the Absent Parameters, the Stars Are Classified into Four Types}
        \label{tab:types of stars}
        \begin{tabular}{lccc} 
    		\hline
    		Type & $T_{\rm eff}$ & $G_{\rm mag}$ & Count\\
    		\hline
    		I & Known & Known & 2147\\
    		II & Unknown & Known & 85\\
    		III & Known & Unknown & 2\\
    		IV & Unknown & Unknown & 4\\
    		\hline
        \end{tabular}
    \begin{tablenotes}
        \item {\textbf{Note.} The counts are collected after the identification of WDs and cool stars (see text), so they add up to 2238 rather than 2575 stars.}
    \end{tablenotes}
\end{threeparttable}
\end{center}

Besides, we assume that the star has blackbody radiation. According to the Stefan-Boltzmann law, we can use the following formula to build up a relation between the $T_{\rm eff}$, luminosity, and radius of the star,

\begin{equation}
    \frac{T_{eff}}{T_0}=(\frac{L}{L_{\odot}}\frac{R_{\odot}^2}{R^2})^{1/4},
    \label{eq:SB}
\end{equation}
where $T_0$=5780K is the effective temperature of the Sun. 

We classified the stars into four categories when completing the parameters (see Table~\ref{tab:types of stars}) and derived the parameters in different ways.

\begin{description}
    \item Type I. For stars with known $T_{\rm eff}$ and $G_{\rm mag}$, we firstly estimate radius via Planck's law. If the radius is below 0.015 $R_{\odot}$, we consider it as a WD and flag it as "WD"; if not, we then estimate the luminosity of the star with Equation (\ref{eq:SB}), and finally obtain the mass of the star both by Equations (\ref{eq:LMR}) and (\ref{eq:RMR}). Note that stars with all parameters known are also classified as type I.
    \item Type II. For stars with known $G_{\rm mag}$, similarly, we similarly derive the radius from $G_{\rm mag}$ and distance, figuring out WDs according to the result. Then we estimate the $T_{\rm eff}$ of the star via $G_{\rm mag}$ and distance, and the remaining steps are the same for type I;
    \item Type III. If the $G_{\rm mag}$ is unknown but $T_{\rm eff}$ is known, we directly estimate the remaining parameters with Equations (\ref{eq:SB})-(\ref{eq:RMR}). This type only has two stars with known \emph{M}, \emph{R} and $T_{\rm eff}$, so we only calculate their luminosity via Eqn.~\ref{eq:SB}.
    \item Type IV. If both of $T_{\rm eff}$ and $G_{\rm mag}$ are unknown, we leave the parameters blank.
\end{description}

\begin{figure*}
    \centering
    \includegraphics[width=0.9\textwidth,height=0.9\textwidth]{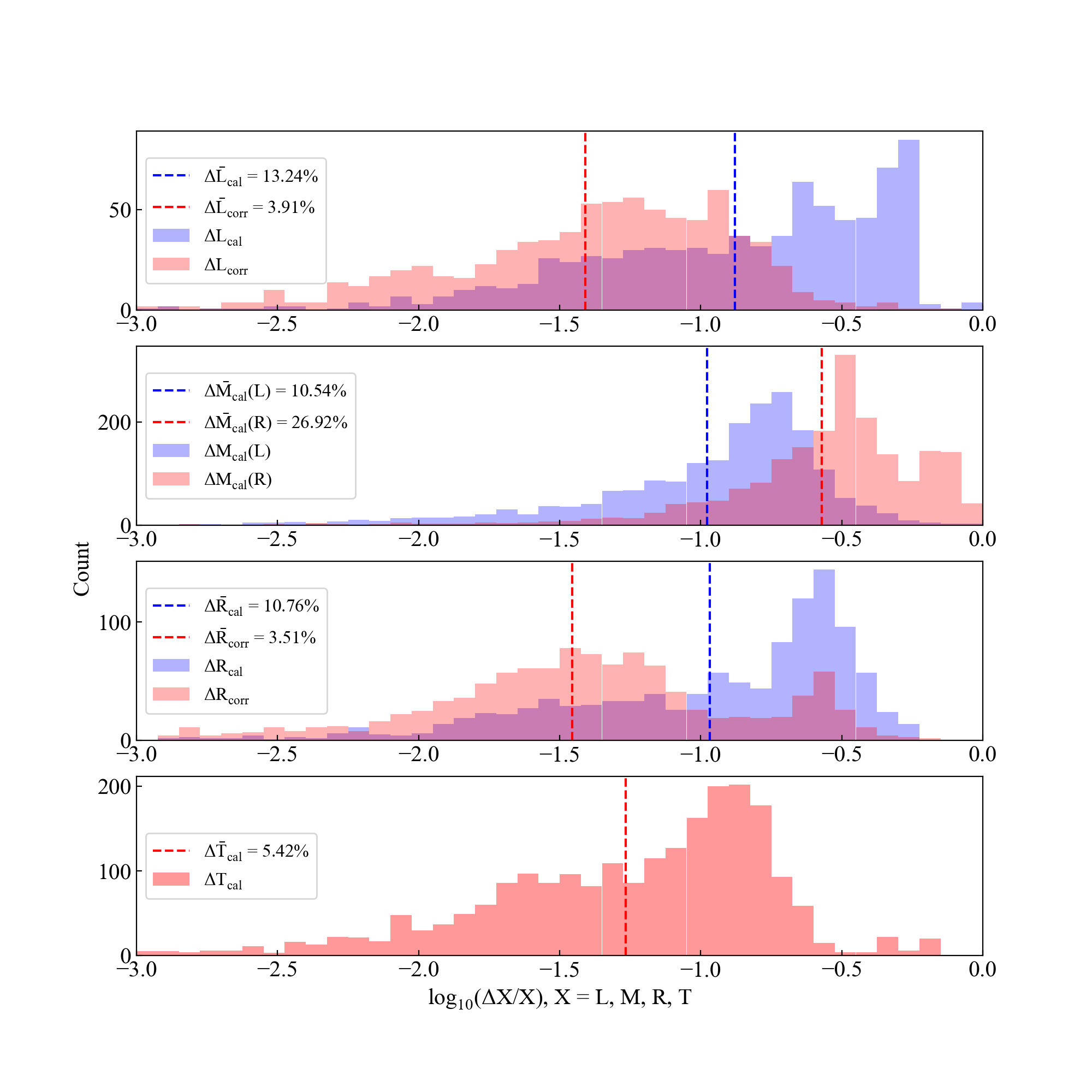}
    \caption{Histogram of the relative error ($\rm \Delta X/X=|X_{\rm cal}-X_{\rm obs}|/X_{\rm obs})$) of the four basic parameters. The dashed lines represent the mean errors of different parameters. The red and blue histograms show the relative errors before and after the correction of the derived parameters, respectively, except for $M_{\rm cal}$ (second panel), which shows the relative errors of stellar mass estimated via corrected luminosity (blue) and radius (red).}
    \label{fig:fit}
\end{figure*}

After deriving the stellar parameters, we compare the derived values of $L$, $M$, $R$, and $T_{\rm eff}$ for stars with known parameters as shown in Fig.~\ref{fig:fit}. We corrected the derived parameters on the basis of the fit result of $L-\Delta L$, which can be seen in Fig.~\ref{fig:fit}. Firstly, we fit the $L-\Delta L$ relation with a piecewise function (Equation (\ref{eq:fit})), and use this function as the correction term for luminosity correction. Here we use $\epsilon_1$, $\epsilon_2$, and $\epsilon_3$ to represent the correction factors for simplicity, which are different fitted functions of the observed $L$. These factors can be checked in our data release on GitHub (\href{https://github.com/IcyDawn/Nearby_Star_Catalog}{GitHub-NSC\footnote{\href{https://github.com/IcyDawn/Nearby\_Star\_Catalog}{https://github.com/IcyDawn/Nearby\_Star\_Catalog}}}). Secondly, we assume that the $T_{\rm eff}$ of a star is unchanged after correction of $L$, and use Equation (\ref{eq:SB}) to calculate the correction factor for the radius. Finally, we compare the accuracy of mass $M$ via $L$ and $R$, finding that the former is better, so we adopt the $L-$estimated mass as $M_{\rm cal}$. From Fig.~\ref{fig:fit}, it is obvious that after correction, $\Delta \bar{L}$, $\Delta \bar{M}(L)$, and $\Delta \bar{R}$ are all around or below 10\%, which is acceptable in this paper. Note that stars with known $L$, $M$, $R$ or $T_{eff}$ are adopted directly, rather than using derived values, and we have not corrected the derived $T_{\rm eff}$ because the relative error is small ($<$10\%):

\begin{equation}
    L_{corr} \cong \left\{
    \begin{aligned}
          & L_{cal}+\epsilon_1 L_{obs}/L_{\odot}, & L_{obs}/L_{\odot}\leq0.01\\
          & L_{cal}+\epsilon_2 L_{obs}/L_{\odot}, &  0.01\leq L_{obs}/L_{\odot}\leq 0.5 \\
          & L_{cal}+\epsilon_3 L_{obs}/L_{\odot}, &  L_{obs}/L_{\odot}\geq0.5 
    \end{aligned}
    \right.
    \label{eq:fit}
\end{equation}

\begin{figure}
\centering
\includegraphics[width=0.9\textwidth,height=0.45\textwidth]{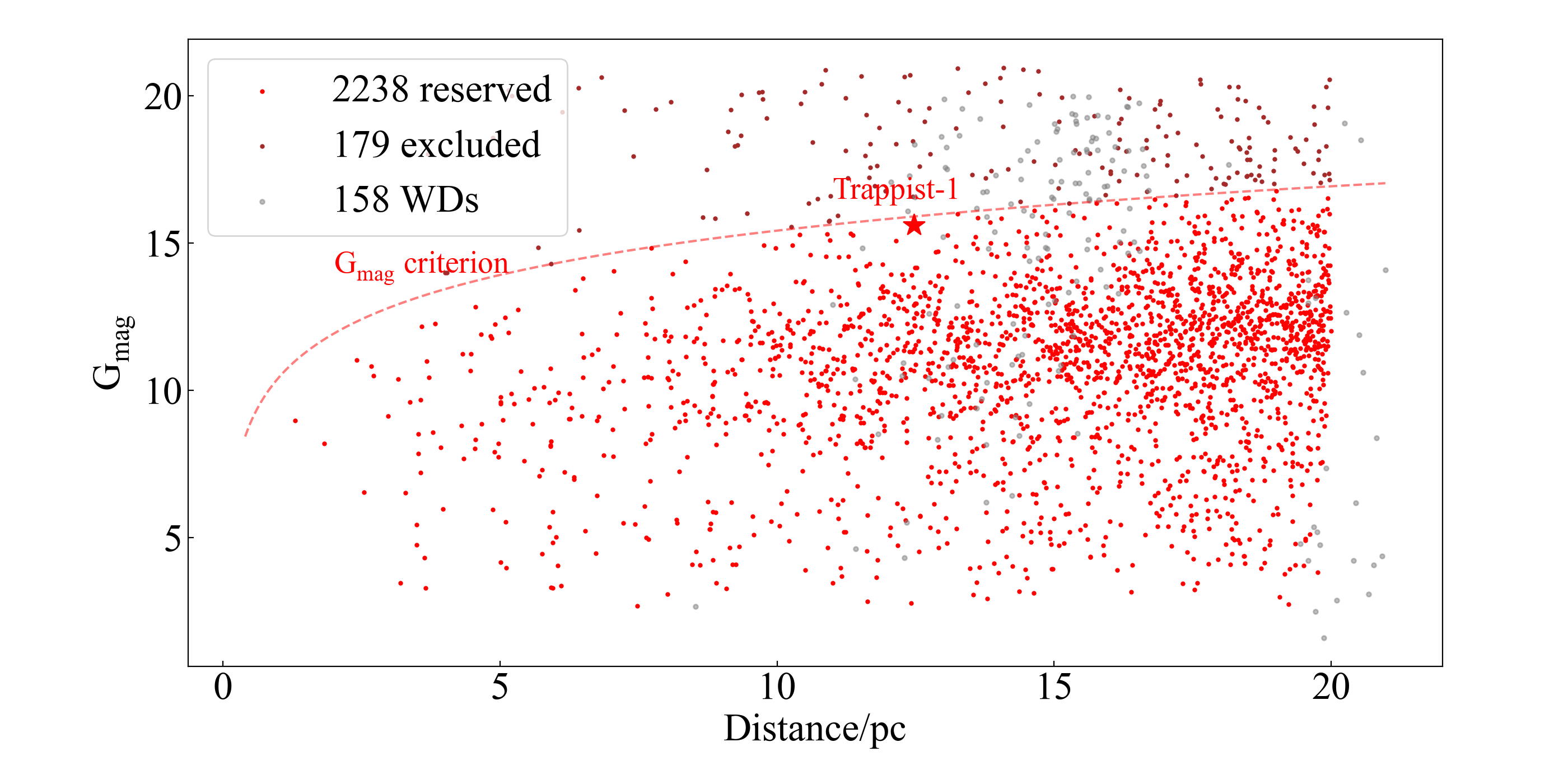}
\caption{Distance and magnitude of all 2575 stars in the NSC. The red line represents the $G_{\rm mag}$ of an M8V star like TRAPPIST-1 (0.1 $R_{\odot}$, 2600K). The $G_{\rm mag}$ of TRAPPIST-1 is represented with a red star. We exclude objects above the red line. Finally, 179 faint objects are excluded. The WDs are not excluded by this figure.}
\label{fig:src_check}
\end{figure}

Finally, the parameters of 2410 stars in NSC are completed, except for 158 WDs, three brown dwarfs, and four stars without any parameters (type IV). After estimating the parameters of different types, we found that some stars have extremely small effective temperatures, masses, or radii. According to TRAPPIST-1, which is known as the coolest star harboring planets, we derive a magnitude-distance relation of an M dwarf with 0.1 $R_{\odot}$ and an effective temperature of 2600K, in the $G$-band, to exclude sources that are fainter than the red dashed line as shown in Figure \ref{fig:src_check}. This step excludes 176 very faint objects, which are probably cool stars ($<$2600 K) or even brown dwarfs (labeled as "BDs"). Finally, 2234 stars are reserved for further estimation of the HZ. The catalog with the completed parameters is defined as the NSC, distinguished from the GCNS.

\begin{figure*}
\centering
\includegraphics[width=1\textwidth,height=0.4\textwidth]{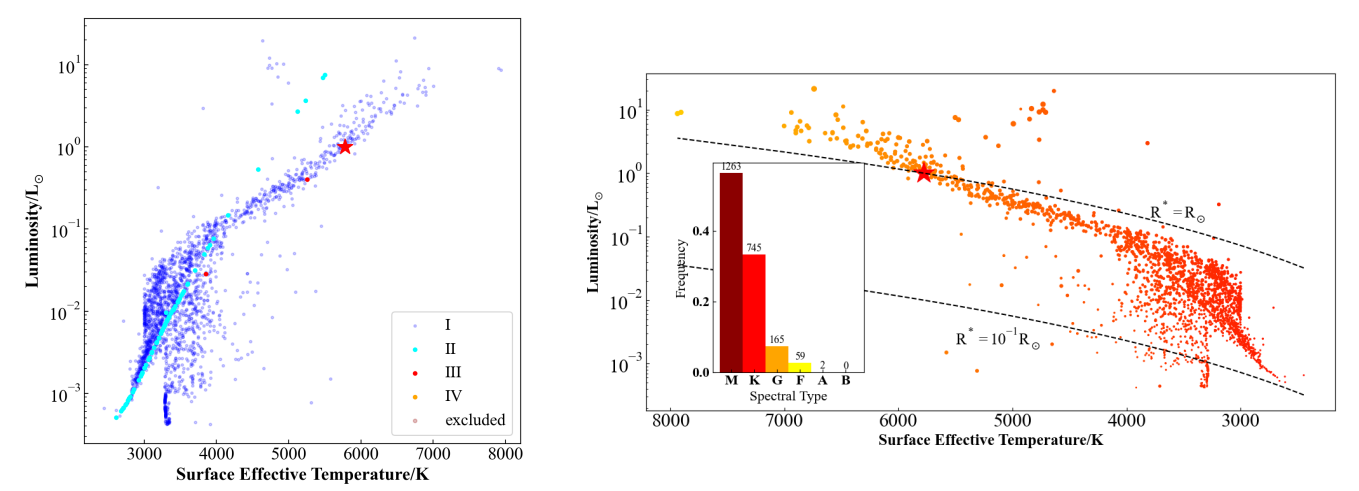}
\caption{The H-R diagram of 2234 stars used to estimate the HZs. Note that this figure does not contain WDs as they are already excluded and have no parameters. In the left panel, the blue points are stars with known $G_{\rm mag}$ and $T_{\rm eff}$, while points with an obvious linear feature are stars whose $L$ \& $T$ are estimated. The brown points are faint objects, which are excluded and are probably cooler stars or even brown dwarfs.}
\label{fig:LTcheck}
\end{figure*}

The H-R diagram of five different types (including excluded cool stars) is shown in the left panel of Figure \ref{fig:LTcheck}. The line structures of the type II and excluded cool stars are because their luminosity is estimated from Eqn.~\ref{eq:SB} with distance and $G_{\rm mag}$. We use Eqn.~\ref{eq:LMR} and \ref{eq:RMR} when estimating mass through luminosity and radius, respectively.

The reserved 2234 stars are suitable for the study of habitable planet detection simulations. We denote the selected subsample as the refined NSC. The H-R diagram of the refined NSC is shown in the right panel of Figure \ref{fig:LTcheck}.

\subsection{Features of Selected Nearby Stars}\label{sec:features}

In this section, we demonstrate the features of 2234 nearby stars in the refined NSC, including their distribution of critical physical parameters and spectral types.

\begin{figure*}
\centering
\includegraphics[width=0.9\textwidth,height=0.45\textwidth]{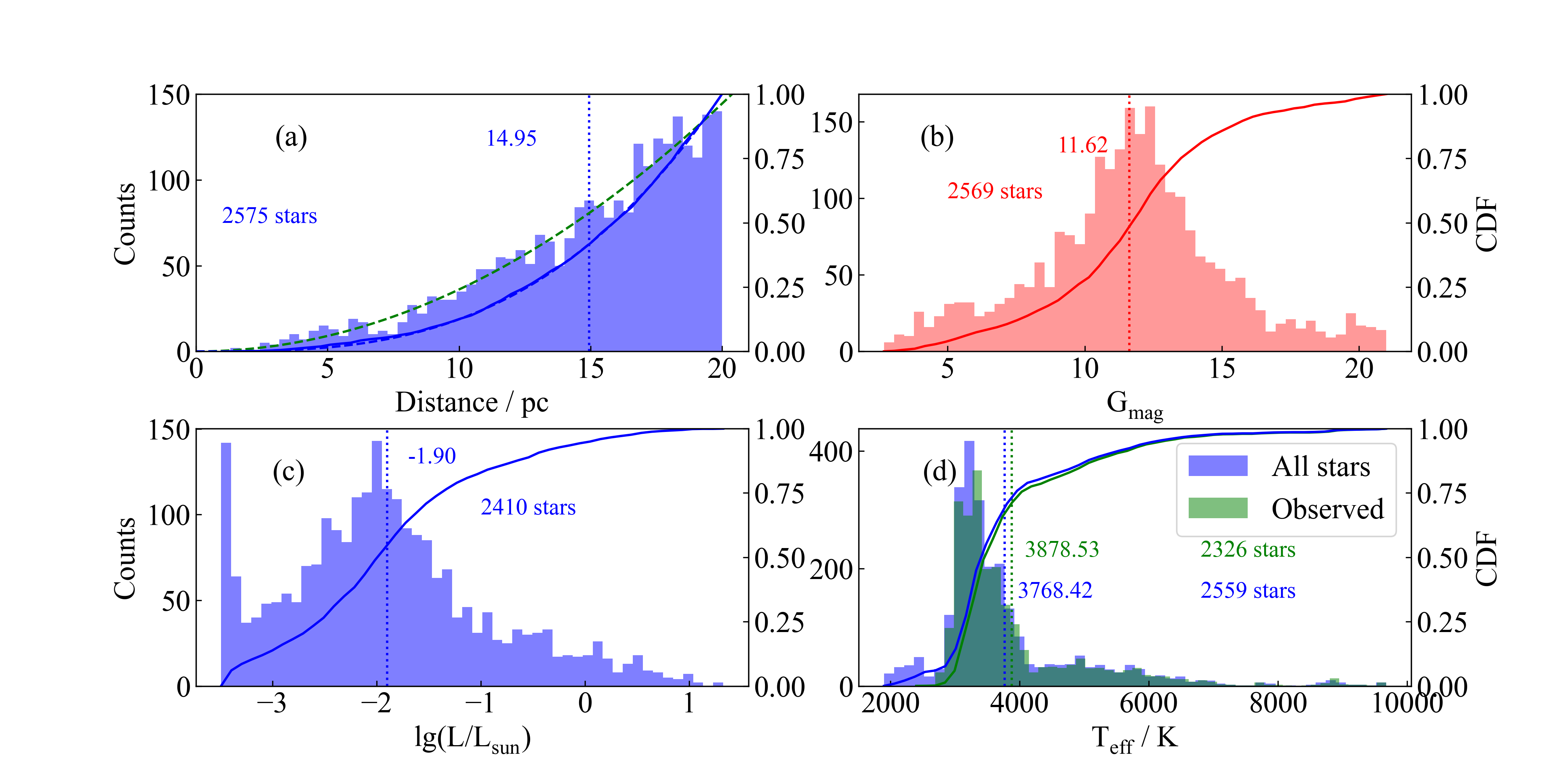}
\caption{Histograms and the Cumulative Distribution Function (CDF) of the (a) distance, (b) $G_{\rm mag}$, (c) luminosity, and (d) $T_{\rm eff}$ of nearby stars. Vertical dotted lines show the mean value of each distribution. Solid lines represent the CDFs. The green and blue dashed curves in panel (a) represent the distribution and the CDF of a uniform spatial distribution of stars, respectively.}
\label{fig:parameters}
\end{figure*}

\begin{figure*}
\centering
\includegraphics[width=1\textwidth,height=0.5\textwidth]{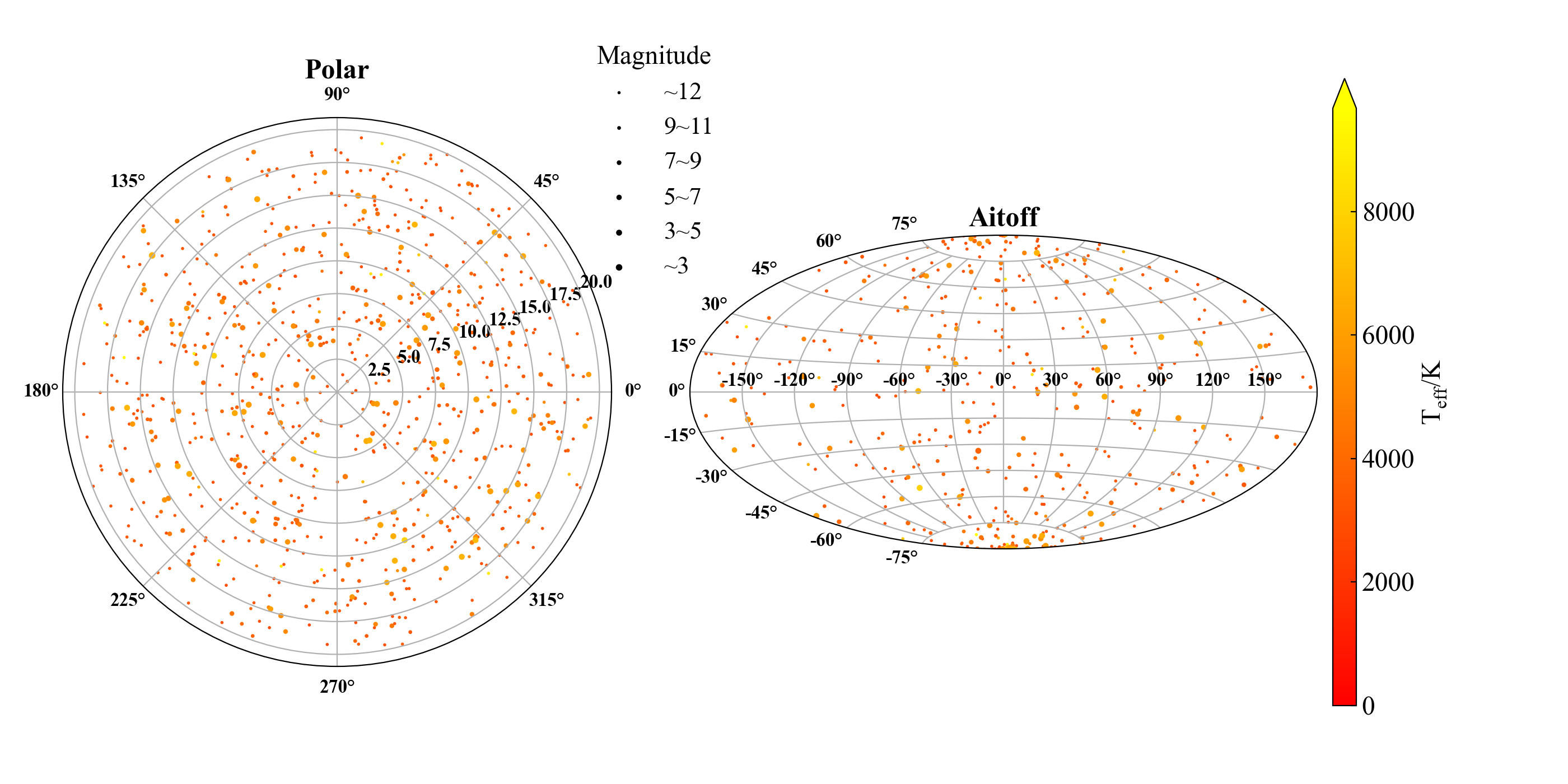}
\caption{The left and right panels are heliocentric polar and Aitoff projections of the refined NSC, respectively. The size of the points represents $G_{\rm mag}$ or $V_{\rm mag}$ and their color represents effective temperature. We can see that stars within 20 pc have an almost spatially uniform distribution.}
\label{fig:distribution}
\end{figure*}

Figure \ref{fig:parameters} shows the distance, magnitude, luminosity, and effective temperature of stars in the GCNS with known parameters, including 2234 stars in the refined NSC. Some parameters are estimated rather than directly observed and are shown in different colors.
In panel (a), the blue dashed and solid lines are consistent, from which we can infer that nearby stars are uniformly distributed in space. Figure \ref{fig:distribution} shows the location of these stars in the sky. Panel (b) shows the magnitude in the $G$ band, and apparently, nearby stars have a peak around $G_{\rm mag}=12$. About 90\% of the stars are fainter than the Sun because 90\% of the stars have a luminosity of less than 1.0 $L_{\rm Sun}$ as seen in panel (c). In panel (d), about 90\% of the stars are cooler than the Sun, which means the majority of stars within 20 pc are red dwarfs in the main sequence.

\begin{figure}
\centering
\includegraphics[width=0.9\textwidth,height=0.45\textwidth]{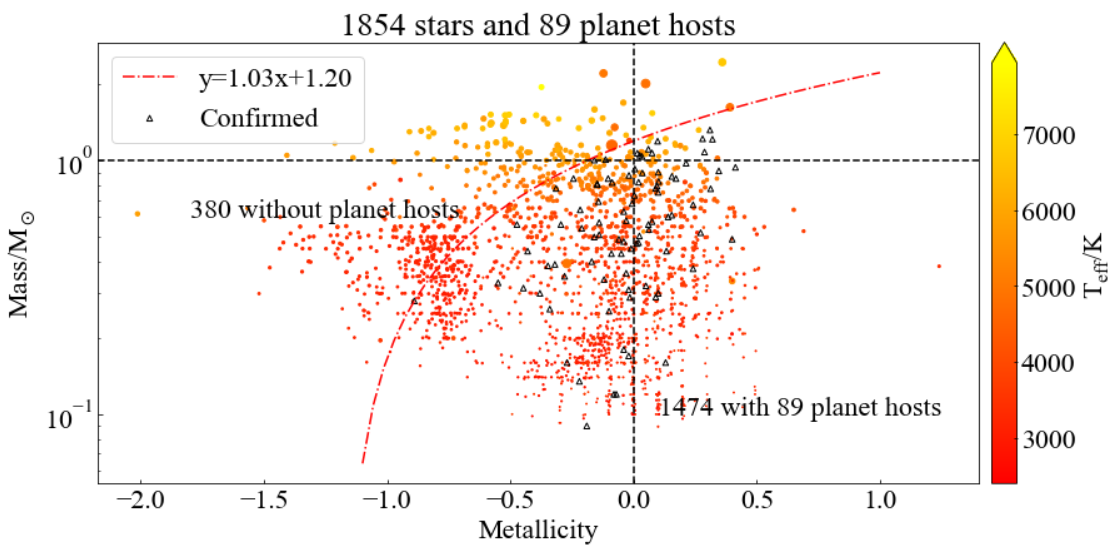}
\caption{Mass-metallicity diagram of 1854 stars with known metallicity. The red curve divides the stars into two populations. Below the curve, there are 1474 stars, including 89 planet hosts ($\sim6.0\%$). Above the curve, 380 stars are left without planet hosts.}
\label{fig:mm}
\end{figure}

At a distance of 20 pc, 145 stars have been confirmed to have 251 planet(s) so far, and 89 of these stars have mass and metallicity measurements. We plot the mass and metallicity of these stars in Figure \ref{fig:mm}. Taking metallicity into consideration, all the 89 stars confirmed to have planet(s) are under the red line. That is, the detected planets prefer hosts with smaller masses and moderate metallicity. So, if we can detect planets around these metal-poor or massive stars, it will probably improve our knowledge of planet formation.

\section{Setting Earth-like planets in the HZ around nearby stars}\label{sec:HZsetting}
To estimate the signal of nearby Earth-like planets via different methods, it is crucial to know the HZ around nearby stars and the location of Earth-like planets. In this section, we introduce how to calculate the inner and outer boundaries of the HZ for the selected 2234 stars and how we set an Earth-like planet in the HZ around each star. 

\subsection{Consideration of Stellar Multiplicity}

Since the stellar multiplicity will lead to instability \citep{1999AJ....117..621H},  the stable region should be considered before we set Earth-like planets around nearby stars. The planet must be in the HZ and be stable in binary systems. Thus, the probability of a stable terrestrial planet in the HZ depends on the overlapped region of the HZ and stable regions. In this section, we try to use Gaia data, especially the RUWE index, to consider the probability of nearby stars, which can be single or in binary systems.

First, we cross-match the NSC with the catalog derived by \citealt{2022MNRAS.513.5270P}, which contains 22,699 binaries within 100 pc in the GCNS, and find 90 matches. All the 90 binaries have estimated separations of $\Delta a$. Later we will use $\Delta a$ we acquire to estimate the stable regions around binaries. Stars that cannot match any sources in the binary catalog could also be binaries. Gaia provides the RUWE index to hint if a star is in binary systems. That is, stars with a large RUWE are probably binaries, while stars with a small RUWE are more likely to be single stars. As shown in Figure~\ref{fig:LTcheck}, $\sim56.5\%$ of the 2234 stars we used to estimate the HZs are M dwarfs, and $\sim90\%$ are M/K stars. According to \citealt{2023hsa..conf..166C}, the multiplicity rate in M dwarfs can be around 40\%. Therefore, for stars with moderate RUWE, we set the probability of being binary stars as 40\%.

Specifically, we divide 2234 sources into three groups according to their RUWE, to consider the binarity.

\begin{description}
    \item (i) RUWE $\leq 1.4$. These 1363 sources are expected to be single stars, and we treat them as single stars; i.e. we do not consider the stable zones.
    \item (ii) $1.4\leq$ RUWE $\leq1.9$ or RUWE unknown. The upper limit of 1.9 comes from the lowest RUWE in the binary catalog \citep{2022MNRAS.513.5270P}. These 569 sources can be binaries or single stars with a probability of 40\% and 60\%, respectively.
    \item (iii) RUWE $\geq1.9$. These 302 stars are likely to be binaries, and we treat them as binary stars.
\end{description}

To calculate the stable zone in binary systems, we follow the previous work by \cite{1999AJ....117..621H}. Generally, there are two types of planetary orbits: S-type and P-type. For different types, the stable zones can be calculated via Eq.(6) and (9) in a previous paper \citep{2021MNRAS.507.4507B}, which depend on the orbital eccentricity, semi-major axis, and the mass ratio of the binary. Based on the derived HZs of 2234 sources, we can add the stable zone into consideration. For a given combination of $(e,q,P)$, we can estimate the semi-major axis of the binary's orbit, which we use to infer the type of the binary. Then we can calculate the stable zone in such a binary system, and, by comparing it with the HZ, the probability of a stable Earth in the HZ can be calculated as the ratio of overlapped area to the area of the HZ. Exempli gratia, if there is no overlapped region between the stable zone and the HZ, the probability is zero, while if the HZ are all inside the stable zone, the probability is 1.

Considering the probability distribution of the binary parameters, we can estimate the weighted average probability $P_{\rm STB}$ of a stable Earth-like planet in the HZ around each possible binary star via calculating a triple integral. To set the probability distribution of the eccentricity $e$, mass ratio $q$, and period $P$(days) of the binary system, we refer to some previous work. Exempli gratia,


\begin{equation}
    f(e) = 1, 0 \leq e < 1,
\end{equation}

according to \citealt{2018MNRAS.474.4322M};

\begin{equation}
    f(q) = \left\{
    \begin{aligned}
          & c_1, & q = M_{secondary}/M_{primary}<0.8 \\\
          & c_2, & 0.8\leq q\leq1.0,
    \end{aligned}
    \right.
\end{equation}

according to \citealt{2004ASPC..318..166D}; and

\begin{equation}
    f(lg P)=8.39\cdot exp(-(lg P-4.8)^2/4.6), 3 
 days\leq P \leq 3000 days,
\end{equation}

according to \citealt{1991A&A...248..485D} and \citealt{2010ApJS..190....1R}.

Note that we truncate the binary period $P$ from 3 days to 3000 days because both the very close and well-separated binaries are similar to single stars, which have a very wide stable zone, including the whole extended HZ. For the cross-matched 90 binaries, we adopt the semi-major axis as $\Delta a$ and fix the period $P$ estimated via stellar mass and $\Delta a$, instead of the probability distribution of $P$. Moreover, the expectation of $a_{\rm STB}$, the semi-major axis of the stable planet in the HZ, can be calculated by the weighted average center of the stable zone inside the HZ.

After we calculate the chances for both S- and P-type binaries to have habitable planets settled in their stable regions, we find that P-type binaries have less chance than S-type binaries. When we estimate the signals of binaries, for simplicity, we ignore P-type parameters in integral calculations. This will only reduce the total expected count of habitable planets by about 40, which is $\sim 2\%$ of all of the planets we considered. After considering the binary stars, there are 123 stars in the binary system that cannot tolerate a stable Earth in the HZ. Thus, the total star sample becomes 2111 stars.

\subsection{Calculation of an HZ}\label{sec:HZcalc}
An HZ is rarely determined precisely because it depends on not only on the stellar spectrum and stellar activity, but also on the planetary atmosphere. In this paper, to calculate the boundary of the HZ around each star, we adopt Equation (2) in \citet{Kopparapu_2013}, which considers the atmospheric effect of different types of stars. We use an extended HZ to enhance the occurrence rate of planets in the HZ. That is, the inner boundary of the HZ (hereafter IHZ) is set to "Recent Venus," while the outer boundary of the HZ (hereafter OHZ) is set as "Early Mars."



\begin{figure}
\centering
\includegraphics[width=0.9\textwidth,height=0.45\textwidth]{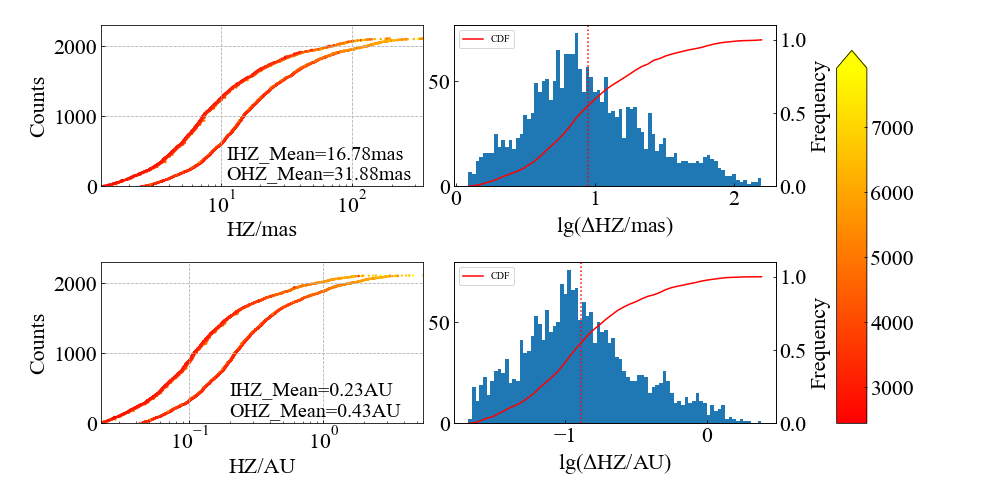}
\caption{Left: location of the IHZ and OHZ. Right: the width of the HZ $\Delta$HZ, for all the 2111 stars. The color bar shows the effective temperature of the stars.}
\label{fig:IHZ and OHZ}
\end{figure}

Figure~\ref{fig:IHZ and OHZ} shows the HZ of all the 2111 stars. Ninety-four stars have an IHZ beyond Earth's orbit (1 au), while 229 stars have OHZ outside 1 au. The average location of IHZ and OHZ is 0.22 and 0.43 au, respectively. Compared with solar systems, the HZs of most stars are closer, since M stars dominate the population. We can also easily obtain the angular extent of the HZs around the nearby stars according to their distance. Forty-one of those stars have IHZs farther than 100 mas, while for OHZs, the number is 157; 130 and 186 stars have IHZs and OHZs ranging from 50 to 100 mas, respectively. The IHZ and OHZ of the stars are mostly settled in the range of 1-50 mas. The average IHZ and OHZ are 16.78 and 31.88 mas, respectively.

Both the median and the mean show that the Sun has a relatively farther the HZ compared to nearby stars. Additionally, the width of the HZ peaks around 0.2 au, which is also much narrower than solar systems because the M stars which are fainter and cooler are the most widely distributed within 20 pc. However, dozens of Earth-like planets in the HZ are detected around M stars due to the observational selection bias.

\subsection{Earth-like planet setting}\label{sec:Earthset}

Since we adopt the extended HZ, we assume an occurrence rate of Earth-like planets in the HZ of 1, and put one Earth-like planet in the HZ around every single star. For binary stars and potential binary stars, the occurrence rate is set as $P_{\rm STB}$. The planets are all fixed to 1 $M_{\oplus}$ and 1 $R_{\oplus}$, since we assume that the planet's atmosphere is similar to Earth; i.e. their emissions and albedo are the same as those of Earth when we put the planet in the HZ around different stars. 

Obviously, not every star has planets in its HZ. The assumption seems optimistic because it does not consider the occurrence rate of Earth-like planets in the HZ, which is correlated with stellar parameters and is not precisely determined yet. In other aspects, some planetary systems are compact, and more than one planet exists in the HZ, e.g., the TRAPPIST-1 system. Therefore, to a certain extent, it is acceptable to assume one planet in the HZ around single stars. The assumption can be altered if we know the occurrence rate of terrestrial planets in the HZ via future surveys of nearby exoplanets.

The location setting is slightly different when we estimate the typical signals of the planet via different methods. In brief, we set the location on the inner and outer boundary of the HZ as two typical places to simulate the signals in section \ref{sec:Signal}. To estimate the detection probability of different methods, we locate the planet in the center of the HZ (CHZ), as described in detail in section \ref{sec:Detect}.

We assume that the orbits of the injecting planets are circles in Sections \ref{sec:Signal} and \ref{sec:Detect}, and when we calculate the maximum signals, we set the preferred inclination that maximizes planet signals.

\section{Signals for different detecting methods}\label{sec:Signal}
In this section, we estimate the signal due to the Earth-like planet in the HZ we set in section \ref{sec:HZsetting}. Different observation methods provide different observational parameters; therefore, we estimate different parameters as the observable signal of exoplanets. I.e., The amplitude of the RV of the planet host is adopted as the observable signal for RV methods. The transit probability (TP), transit duration (TD), and transit depth (Dep) are adopted as the typical signals for transit methods. The maximum stellar displacement can be adopted as the typical signal for astrometry methods. For direct imaging methods, the maximum angular separation is used to estimate the resolution requirements, and the radiation intensity ratio is the typical signal. More details can be found in the following subsections.




\subsection{The RV method}\label{sec:RVest}
The RV has been a popular method used to find exoplanets since the early stages of the exoplanet era. In order to estimate the amplitude of the RV of the host stars induced by their planets, we use the Eq 46 of \citet{clubb2008detailed}, i.e. the following formula:

\begin{equation}
    V_r= \frac{{2\pi}a_p}{P\sqrt{1-e^2}}\frac{M_{p}\sin{i}}{M_{star}}(cos({\omega}+f)+e\cdot cos{\omega})
\end{equation}
where $V_r$ is the RV of the host stars, $P$ is the period of the planet, $a_p$ is the semi-major axis, $i$ is the inclination, $e$ is the eccentricity, $f$ is the true anomaly, and $\omega$ is the argument of perihelion. 

To calculate the RV semiamplitude of the star, we use Eq 57 of \citet{clubb2008detailed}. We set the eccentricities of the planets as zero. Thus, the semiamplitude $K$ of the RV can be simplified as follows: 










\begin{equation}
    K=0.09\sin{i}(\frac{M_p}{M_
\oplus})\sqrt{(\frac{1 AU}{a_{p}}) (\frac{M_{\odot}}{M_{star}})}  m/s. 
\label{Equation RV}
\end{equation} 

In Eq \ref{Equation RV}, $0.09$ m/s is the amplitude of the RV of the Sun caused by the Earth. Furthermore, the amplitude depends on the inclination of planetary systems. For a planet orbit with random orientation, the averaged $\sin{i}$ is $\frac{\pi}{4}$. Thus, we add a factor of $\frac{\pi}{4}$ instead of $\sin{i}$ to model the averaged RV amplitude, i.e. 

\begin{equation}
\bar{K}=0.09 \times \frac{\pi}{4} \times \sqrt{(\frac{1 AU}{a_{p}}) (\frac{M_{\odot}}{M_{star}})} m/s.
\end{equation}


\begin{figure}
\centering
\includegraphics[width=0.9\textwidth,height=0.45\textwidth]{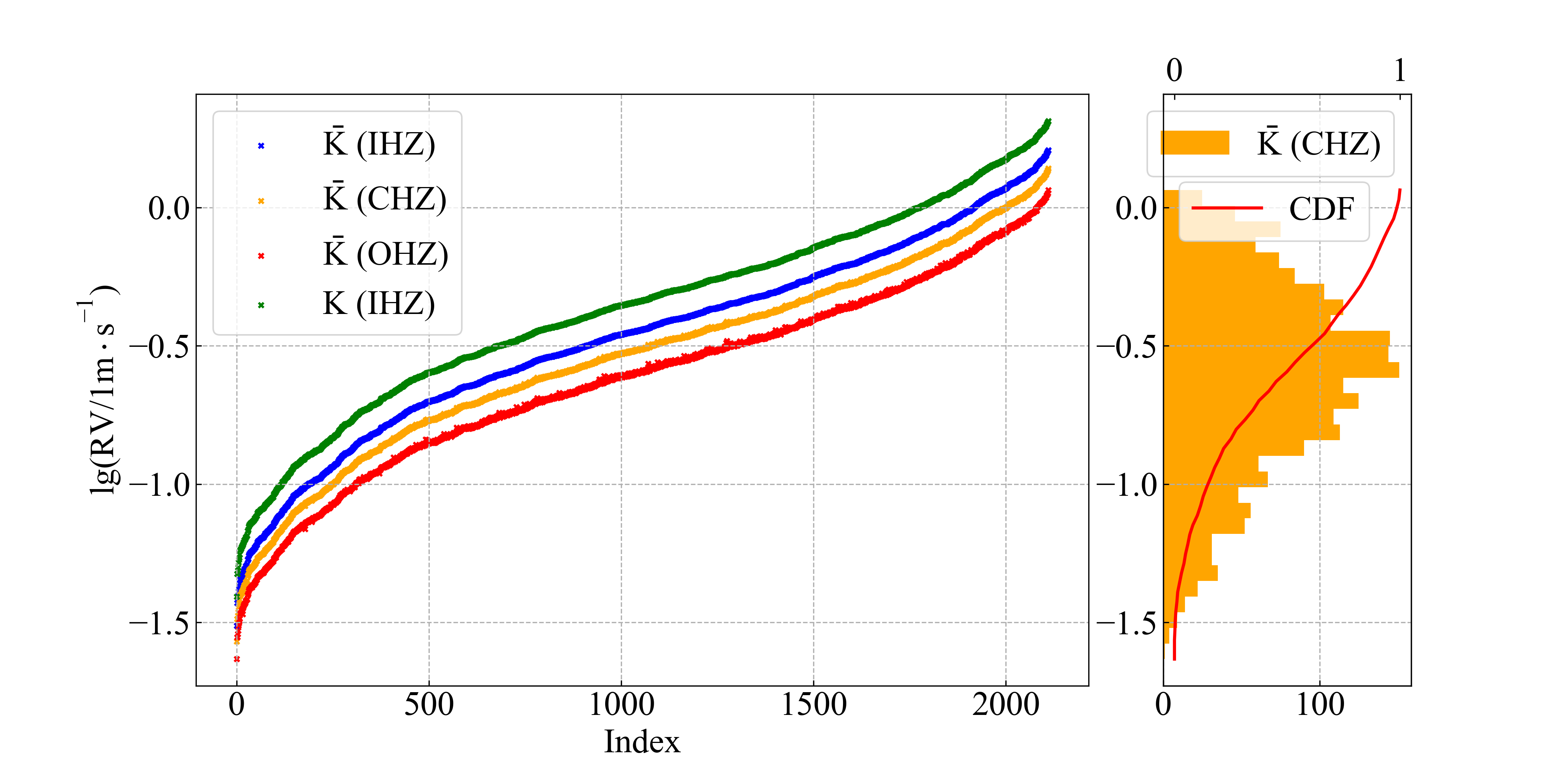}
\caption{The blue, orange, and red crosses represent the averaged RV of stars due to Earth-like planets located at the IHZ, CHZ, and OHZ, respectively. The green crosses represent the maximum RV of Earth-like planets at the IHZ, with an inclination of $90^{\circ}$. The right panel shows the histogram of the RV expectation and the CDF.}
\label{fig:RVplot}
\end{figure}

The inner and outer boundaries are chosen as the typical location of an Earth-like planet. From the formula, the amplitude decreases when the location of the planet varies from the inner boundary to the outer boundary of the HZ. As shown in Figure \ref{fig:RVplot}, we plot the distribution of the averaged amplitude ${\bar{K}}$ when the planets are at the inner and outer boundaries of the HZ. There are 336 stars located at the inner boundary of the HZ with a maximum semiamplitude above 1 m $s^{-1}$, while 144 stars at the outer boundary have a maximum semiamplitude above 1 m $s^{-1}$. Of those located on the inner boundary, 2016 have a stronger maximum amplitude than Earth, while for those on the outer boundary, the number is 1940. What is more, we find that the amplitude caused by the planet on the inner boundary is roughly proportional to the amplitude caused by the planet on the outer boundary. That is, for 1555 stars, the RV amplitudes caused by the planet at the IHZ are $40\%-46\%$ larger than those at the OHZ, while for 1974 stars, the enhancement fraction varies from $35\%$ to $46\%$.

Finally, we demonstrate the histogram of RV amplitude due to the Earth-like planets located at the CHZ, as shown in the right panel of Figure \ref{fig:RVplot}. We can see that the amplitude is concentrated around 0.2 m $s^{-1}$, and there is a small peak around 0.08 m $s^{-1}$; the latter is similar to the amplitude of the Sun caused by Earth. Most Earth-like planets around stars in the refined NSC show a larger RV amplitude than the Earth-Sun system.

\subsection{The transit method}\label{sec:transitest}

As the most fruitful method, the transit method is used via Kepler and TESS to find thousands of exoplanets. The signal of transit is related to the TD, TP, and Dep. All of them are coming from the following equations scaled by the Earth:

\begin{equation}
  \begin{split}
       &TP =0.005 \times(\frac{R_{star}}{R_{\odot}})(\frac{1 AU}{a_p})\\
       &TD = 13 \times \frac{\pi}{4} \times (\frac{R_{star}}{R_{\odot}})\sqrt{(\frac{a_{p}}{1 AU})(\frac{M_{\odot}}{M_{star}})} hour \\
       &Dep={(\frac{R_{p}}{R_{star}})}^2
  \end{split}
\end{equation}
where $R_{p}$ is the radius of the planet, which is set as the radius of the Earth. The factor $\pi$/4 is due to the average effect of impact parameter $b$, which is assumed as a uniform distribution from zero to 1.

\begin{figure}
\centering
\includegraphics[width=0.9\textwidth,height=0.45\textwidth]{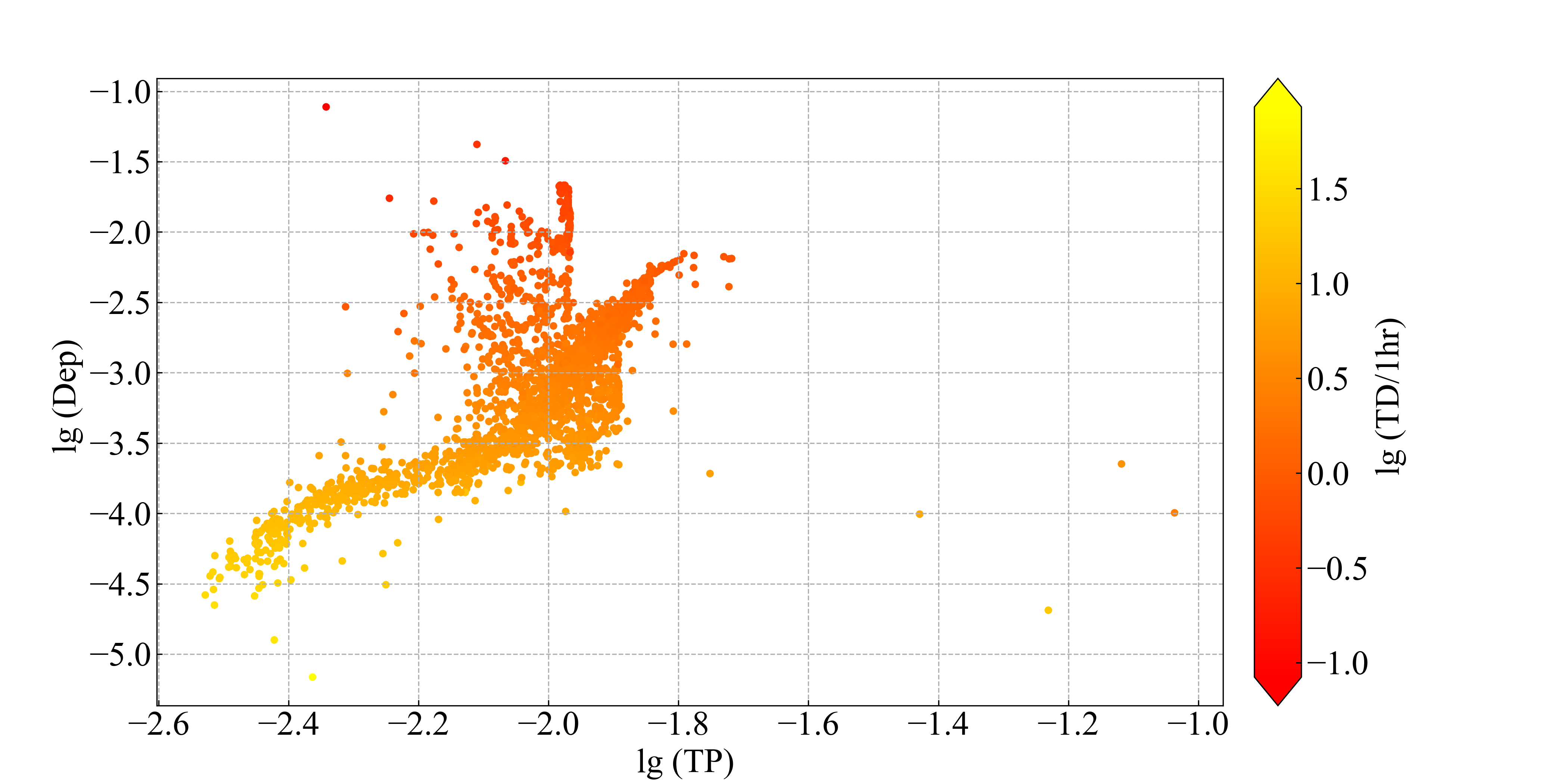}
\caption{The transit depth ($Dep$), probability($TP$) and duration ($TD$) of Earth-like planets located at the center of the HZ.}
\label{fig:Transit}
\end{figure}

The Dep does not change no matter where the Earth-like planet is located, while the $TP$ decreases when the planet becomes further away.
Figure \ref{fig:Transit} shows the Dep, TP, and TD generated by Earth-like planets located on the CHZ, and the average values are 2268ppm, 0.97\%, and 4.64 hr, respectively. Most ($\sim 93.8\%$) planets have a TD shorter than 13 hours (Earth). As the TD increases, both the TP and Dep decrease. Especially the TD and Dep are negatively correlated because the larger stellar radius will lead to a larger TD and smaller Dep. 

Compared with the Earth-Sun system, most planets have a larger Dep, as well as TP, while the TD is shorter. It has to be noticed that the transit method needs a strict geometric configuration. Although the Deps of nearby M stars are larger than the typical values of Earth-Sun, the detection of these planets strongly depends on the TP, which can be seen in section \ref{sec:transitdetect}.

\subsection{The astrometry method}\label{sec:astrometryest}

Astrometry is a method to observe the displacement of a star by its planet in order to find the planet. The maximum displacement $A$ can be estimated by the following equation of the Earth to the Sun:

\begin{equation}
    A=3(\frac{M_p}{M_{\oplus}})(\frac{a_p}{1AU})(\frac{1pc}{d})(\frac{M_{\odot}}{M_{star}}){\mu}as,
    \label{astrometry formula}
\end{equation}
where $d$ is the distance of the star. Compared with other methods, the astrometric signals are different from Equation (\ref{Equation RV}); i.e. larger semi-major axes or closer systems lead to larger displacements. Therefore, the astrometry method prefers finding nearby planets with a longer period. Additionally, since the HZ around massive stars becomes further ($\propto M_{\rm star}^2$ approximately), a planet in the HZ around massive stars leads to larger astrometric signals $A\propto M_{star}$, which is in contrast to the RV and transit methods that prefer planets around M stars.

\begin{figure}
\centering
\includegraphics[width=0.9\textwidth,height=0.45\textwidth]{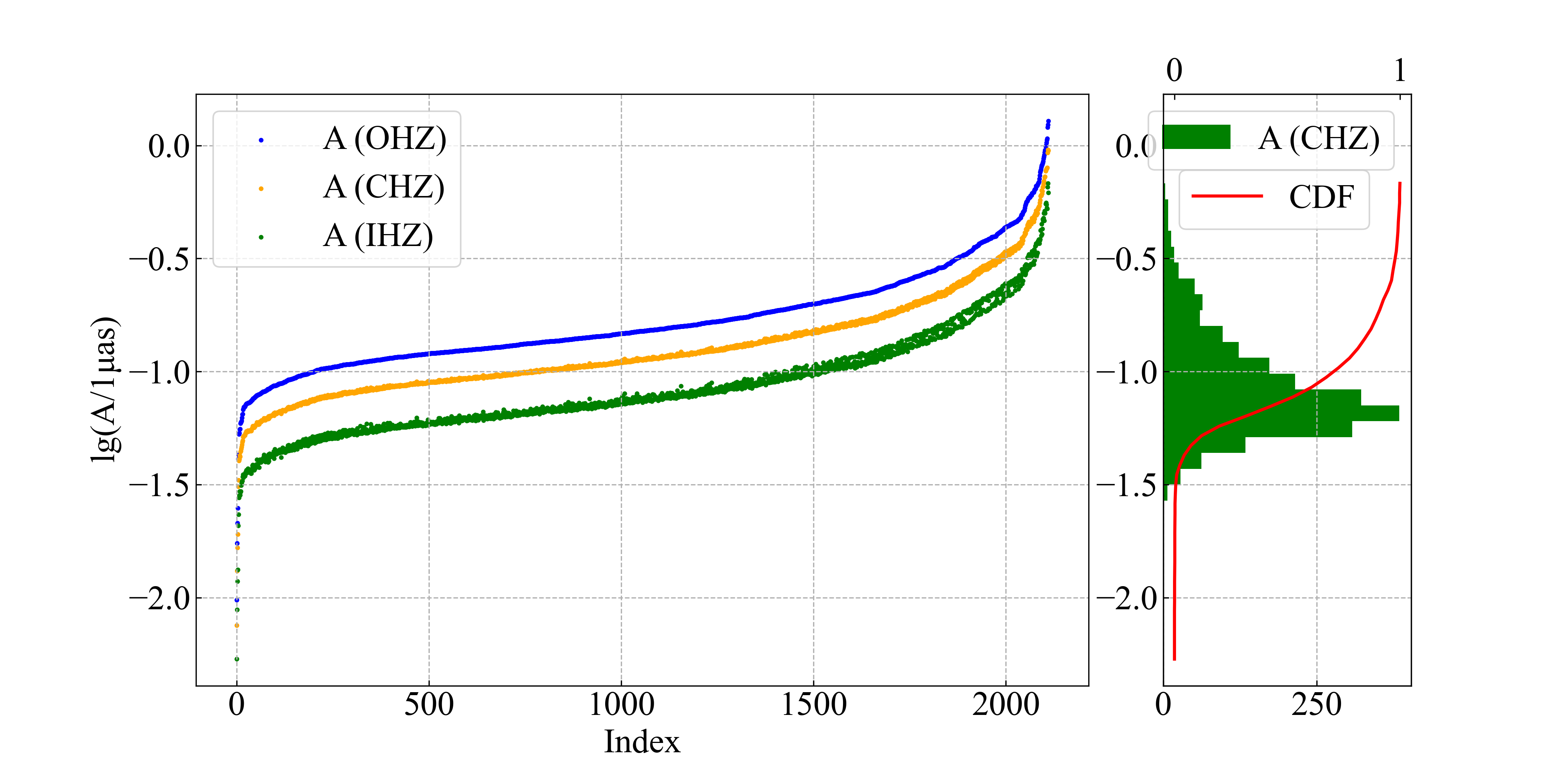}
\caption{Left: astrometric displacement $A$ of stars due to Earth-like planets located at the IHZ, OHZ, and CHZ. Data are sorted by OHZ. Right: The distribution and CDF of the stellar displacement due to planets at the CHZ.}
\label{fig:Astrometry}
\end{figure}

Due to the limits of current astrometric precision, the astrometry can hardly detect Earth-like planets in the HZ with the precision of Gaia, which is above 50$\upmu$as, while the largest calculated signal is only $1.3\upmu$as. According to Figure \ref{fig:Astrometry}, astrometric signals are concentrated around 0.05-0.1 $\upmu$as, because most stars are M dwarfs. Although M stars have smaller masses, the HZ is also closer (see Figure \ref{fig:Astrometry Expectation}). Planets located at the OHZ are nearly twice as far as those at the IHZ; thus, the displacement in the blue line in Figure \ref{fig:Astrometry} is twice that of the red one. Only seven systems have maximum displacements $>1 \upmu$as; thus, submicroarcsecond precision is required to detect most nearby Earth-like planets in the HZ.

\subsection{The direct imaging method}\label{sec:imagingest}




The flux of the planet is contributed by both reflections of stellar light and the thermal radiation of the planet. When the Earth-like planet reflects the light from its host star, a uniform albedo is assumed to be independent of the wavelength and marked as $\alpha$. To calculate the thermal emission, we assume a blackbody spectrum, and the temperature of the planet is uniform; i.e., the dayside and nightside have the same temperature. Absorption and scattering due to clouds are ignored in the following calculation. 
To compare the contrast at different wavelengths, we choose different wavelengths from visible to the near-IR band; i.e. 550 nm represents the visible light, and 10 $\upmu$m represents the infrared wavelength. Also, we assume that the temperature of the planets located in the CHZ is 288 K, which is equal to the average temperature of Earth, to make a rough estimation. The thermal flux can be calculated via the Planck formula. The reflected flux of the planet can be calculated as $\alpha f_{\rm phase}(\frac{r}{a_p})^2 F_{\rm star}$. Here$F_{\rm star}$ is the stellar flux, and the albedo $\alpha$ is set as 0.29, and $f_{\rm phase}$ is a factor that shows how much the surface of the planet facing us is illuminated. When the planet reaches its maximum angular separation from the star, $f_{\rm phase}$ can be set as approximately 0.5. We fixed $f_{\rm phase}=0.5$ in the following calculations.



\begin{figure*}
\centering
\includegraphics[width=1\textwidth,height=0.5\textwidth]{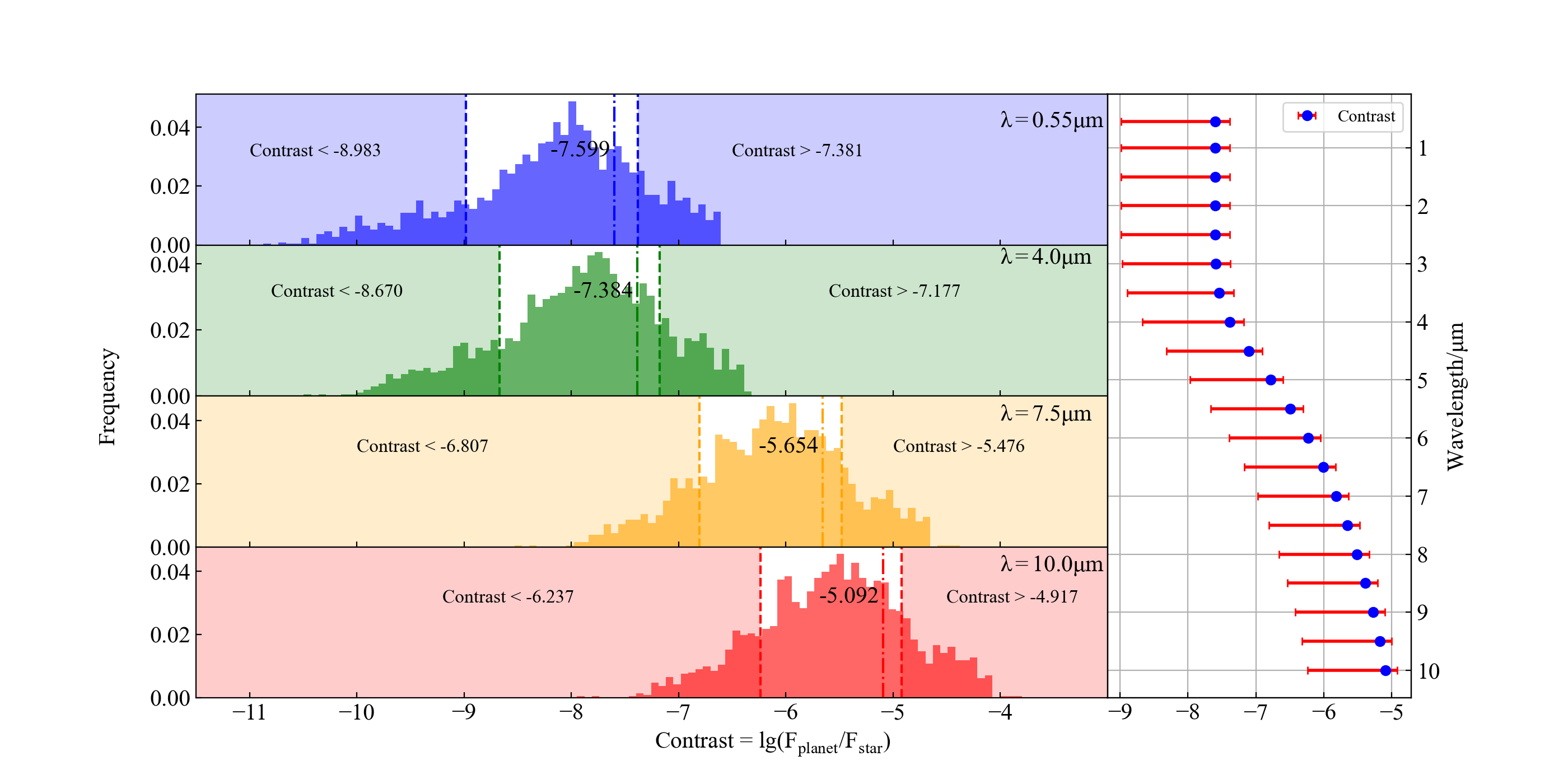}
\caption{Left: contrast between the Earth-like planets on the CHZ and stars at four different wavelengths from 550 nm to 10 $\mu m$. Vertical lines represent the averaged contrast and the $1\sigma$ range. In the right panel, the averaged contrast varies from $10^{-8}$ to $10^{-5}$, while the wavelength changes from $0.55nm$ to 10 $\mu m$.}
\label{fig:flux-ratio}
\end{figure*}

By combining the thermal radiation and the reflected radiation at different wavelengths, we can easily get the total ratio of the radiation of the planets to the radiation of the stars. The contrast between star and planet varies with the wavelength, as shown in Figure \ref{fig:flux-ratio}. The small contrast at the visible wavelength (550 nm) is about $2.51 \times 10^{-8}$. The contrast increases when observed at longer wavelengths, since the blackbody radiation dominates. Especially at 11 $\mu$m, where the radiation of the planet peaks with a temperature of 288 K, the average contrast of all planets becomes $\sim 8.09 \times 10^{-6}$. Also in Figure \ref{fig:flux-ratio}, we show the mean values and standard deviations of the contrast ratio at different wavelengths.

Last, we also need to consider the angular resolution to distinguish the planet from the host stars. As shown in Figure \ref{fig:IHZ and OHZ}, the averaged separation is around 10 mas; thus, according to the Rayleigh criterion, the diameter of a single aperture or the baseline of a synthetic aperture should be larger than $ \frac{1.22\lambda}{10 mas}$. Considering the extreme contrast, both chronograph and nulling interferometry require an inner working angle at least 1-2 times than the Rayleigh limits (see JWST HCI documents\footnote{https://jwst-docs.stsci.edu/methods-and-roadmaps/jwst-high-contrast-imaging}). Thus, for the optical band (500 nm), the aperture diameter needs to be around 30 m, and in the infrared band (10 $\mu$m),  the baseline of nulling interferometry needs to be as long as hundreds of meters.


\subsection{Comparison of signal features via different methods}

Here we will compare the different signal features via the four methods mentioned above. It is obvious that the signals of the RV method and the transit methods have a strong connection with their mass and radius; i.e. the maximum amplitude of the RV and the Dep of Earth-like planets around M stars are larger than those around hotter stars. The signals of both RV and transit are concentrated around values larger than the typical system (Earth-Sun), because of the dominant population of M stars in NSC. For instance, in Figure \ref{fig:RVplot}, we can see two peaks; the higher peak shows the aggregation of signals of M stars, while the lower peak is the aggregation of those of G stars. 

On the contrary, for Earth-like planets around M stars, the average displacement of astrometry is lower than the typical Earth-Sun system. Although the contrast between planets and M stars becomes lower, which benefits us by detecting planets via direct imaging, the smaller angular separation challenges the current instruments.

Due to different detection biases, finding Earth-like planets in the HZ around G stars via astrometry or direct imaging has great potential and can be anticipated in future space missions.

\section{Prediction of Detecting an Earth-like Planet}\label{sec:Detect}

So far, we have detected 252 planets within 20 pc, but according to the data of \citet{2023AJ....165...34H}, only 8 of them are confirmed to be Earth-like planets, which are located in the HZ with a radius between 0.5 and 1.5 $R_{\oplus}$. In this section, we make a rough judgment about how many planets in total we can detect within 20 pc with different observing methods. We assume that the Earth-like planet every star owns is exactly located in the CHZ. For a planetary system, if the observable signals produced by an Earth-like planet with different methods are significant in certain criteria (see each subsection in detail), we will regard it as a detectable planet. We also distinguish different spectral types of stars according to Table 5 of  \cite{Pecaut_2013} to show planets around the kinds of stars that are preferred in each method.


\subsection{RV Methods}\label{sec:RVdetect}

\begin{figure}
\centering
\includegraphics[width=0.45\textwidth,height=0.45\textwidth]{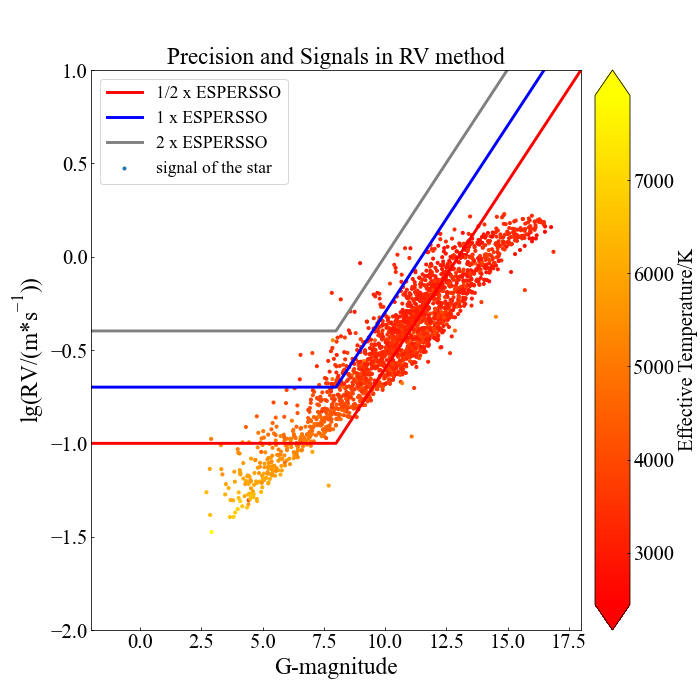}
\caption{Comparison between the maximum RV signal for planets in the CHZ, and three different precision models. The blue line is the model based on ESPRESSO, while the red and gray lines represent precision that is half and twice that of ESPRESSO, respectively.}
\label{fig:RV Expectation}
\end{figure}

Exoplanets within 20 pc are mainly detected by RV; i.e., 221 in 252 exoplanets are detected by RV, while four of them are Earth-like planets. As the most powerful RV instrument, ESPRESSO is expected to have the best accuracy of 0.1 m$s^{-1}$ for the stable stars with $V_{\rm mag}$=6, but in observations of quiet stars, the instruments can reach about 20cm $s^{-1}$ \citep{Netto_2021}. According to the results, we adopt a theoretical precision model based on Poisson noise, as shown in Figure \ref{fig:RV Expectation}. The best precision (single data) of 0.2 m $s^{-1}$ is assumed for stars brighter than $G_{\rm mag}=8$ to get an optimistic estimate. Thus, to determine the criteria of planet detection, we choose a signal-to-noise ratio (S/N)=1 because of the enhancement of S/N via phase folding. Exempli gratia, \citet{2016Natur.536..437A} confirmed Proxima Centauri b by HARPS, a terrestrial planet with RV amplitude of 1.38 m $s^{-1}$. Considering the typical precision of HARPS of $\sim$1 m $s^{-1}$ (single measurement), the S/N is $\sim$0.69, and can be detected by RV after phase folding.

In Figure \ref{fig:RV Expectation}, the blue line corresponds to the RV precision of ESPRESSO, while the red and gray lines represent precision that is half of and twice that of ESPRESSO. Similar to section \ref{sec:transitdetect}, we set the Earth-like planet in the middle of the extended HZ, and the RV signal is calculated as the maximum amplitude in edge-on cases (inclination=0). Figure \ref{fig:RV Expectation} shows that three stars have RV signals larger than the gray lines. Eighty-nine stars have RV signals larger than the precision of ESPRESSO, which has the detectable criterion of S/N$>$1 (blue line). In a more precise criterion model, i.e. 0.1 m $s^{-1}$ at mag=8, there are 1073 stars that have RV signals larger than the criterion.


Since the RV amplitude depends on planetary inclination, assuming the orientation is randomly distributed in the sky, we can estimate the detection probability of a planet with a maximum RV amplitude of $K_{max}$. If the maximum amplitude $K_{max}$ is greater than the precision criterion $K_{c}$, the detection probability of the planet should be $\cos{(\sin^{-1}{(K_{c}/K_{max})})}$. In the case of $K_{max}<K_{SNR=1}$, the probability becomes zero. The total number of detectable planets is obtained by summing up the probabilities of all of the planets. After considering the inclination, with the precision of ESPRESSO (blue line), the total expected number of habitable planets that can be detected is 39. 

When we take star types into consideration, we can clearly see the selection bias of the RV method. That is, 28 of 39 detectable planets are around M-type stars, while the left 10 planets are around K stars, and the remaining one is around a G star. Because M stars usually have a lower mass and closer HZ, the maximum semiamplitude of the RV induced by planets is usually larger than $0.3 m/s$, as seen in Figure \ref{fig:RV Expectation}. For M stars brighter than a $G_{\rm mag}$ of 11, most planets have signals larger than the red line, while for these fainter M stars, the RV signals are hardly detected because of the fainter signals compared with the precision criterion.

Based on the results considering the inclination distribution, we can find that only one planet around 150 G stars, with a detection threshold of $>0.2 $m $s^{-1}$, i.e., the limit of ESPRESSO for bright stars. And if we change the criterion of S/N $\geq 1$ to $\geq \frac{1}{2}$ (red line in Figure \ref{fig:RV Expectation}), which is similar to the RV signals of the Earth-Sun system, nine planets around G stars are expected to be detected. The enhanced number is mainly because of the limitation of precision. Note that more than 90\% of the Earth-like planets around G stars cannot be detected with such precision.

\subsection{Transit Methods}\label{sec:transitdetect}

\begin{figure}
\centering
\includegraphics[width=0.45\textwidth,height=0.45\textwidth]{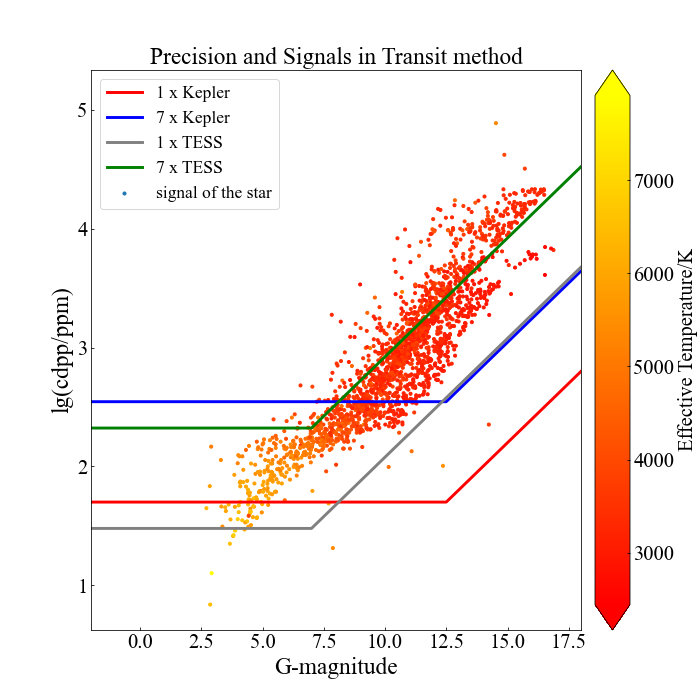}
\caption{Transit signals and two photometric precision models for each instrument's precision. The red line represents the precision of Kepler's CDPP, and the gray line represents TESS's. The blue line and green lines represent noise models seven times lower than Kepler and TESS, respectively. Note that most planets found by Kepler are above the blue line with an S/N $>$7. The color shows the temperature of the stars.}
\label{fig:Transit Expectation}
\end{figure}

Of all those detected planets within 20 pc, 24 of them are first observed by the transit method. Since the planets closer to their host stars are easier to detect via transit, all of them are out of the HZ. That is, they suffer a large amount of radiation from their host stars and become too hot, except for those around cool M stars. The most successful telescope using the transit method to detect planets is Kepler. For stars whose magnitude is 12, the typical noise level is 30 ppm \citep{2018ApJS..235...38T}. But by calculating the mean value of the precision of Kepler DR25, the mean precision is 50 ppm for stars with magnitude $K_p=$12.5 by a 6 hr long observation. On the other hand, TESS, as an all-sky survey mission, has a value of the precision of about 30 ppm for stars with magnitude $T_p=$7 \citep{Oelkers_2018} in the same observing time. Therefore, we choose the above two precision model, i.e. we adopt 50 ppm as the best precision for stars brighter than $G_{\rm mag}$ = 12.5 (the precision of Kepler), or we adopt 30 ppm for stars brighter than $G_{\rm mag}$ = 7 (the precision of TESS) to make the following estimation. Note that the photometric accuracy is dominated by the Poisson noise of the photons, which depends on the magnitude of the host star. The precision model adopted based on Kepler and TESS mission is also shown in Figure \ref{fig:Transit Expectation}.

In Figure \ref{fig:Transit Expectation}, the red line represents the precision of Kepler in a theoretical Poisson noise model. The blue line shows another different precision curve; i.e.; the precision is seven times lower than Kepler (hereafter 7 $\times$ Kepler). Similarly, the gray and green lines show the same precision curves of TESS, respectively. We set the Earth-like planet in the CHZ, then the Dep is also plotted in Figure \ref{fig:Transit Expectation}. Note that most detected Kepler planets have an S/N larger than 7; thus, among all the 2111 planet-star systems, $69.9\%$ of them are above the line of 7 $\times$ Kepler, which means they are detectable via the transit method with the 7 $\times$ Kepler criterion if their orbits have a suitable inclination. Compared with the 7 $\times$ TESS criterion, there are only $28.1\%$ of them have signals above the criterion.

However, the TP depends on both the distance of the planet and the stellar radius, which is usually $<1\%$ as shown in figure \ref{fig:Transit}. To predict the detected number of planets, we sum these planets above the detection criterion, weighted by TP, to take the random orbit orientation into account. The result shows that, in the criteria of 7 $\times$ Kepler and 7 $\times$ TESS, there are 12.5 and 4.9 Earth-like planets in the HZ that are expected to be observed by the transit method, respectively. Comparing the number of signals higher than the criterion and the number of planets that are expected to be detected, we strengthen the concept that the predominant factor that minimizes the expectation is the low transit possibility. 

In the prediction of detectable transit planets, we do not consider the impact parameters, which will definitely influence the shape of the light curves and influence the detectability. Furthermore, a planet with a long period may only have one transit event detected, which can hardly be confirmed as a planet. But it can be confirmed via transmission spectrum follow-up, as the TD is long enough around bright stars. We assume that the observation can be done in a secular time-spanning baseline (e.g. the ET Mission aims to monitor the Kepler field for more than 4 yr, \citealt{2022arXiv220606693G}), thus, a planet with a long TD and long period is also considered detectable, but only if the S/N exceeds the threshold. 

As shown in Figure \ref{fig:Transit Expectation}, with Kepler's precision, the transit method prefers detecting Earth-like planets around M stars (12.11 planets), while they are seldom around G stars, as the expected number is only 0.013. With TESS's precision, the result shows the same trends; the expected numbers are 4.56 and 0.021 for planets around M stars and G stars, respectively. Obviously, M stars with a smaller radius and closer HZ lead to larger depths and TPs. Most planets around M stars have detectable signals with a precision like Kepler; however, the activity of M stars dominates the time domain noise rather than the Poisson noise and will therefore sharply cut down the predicted number of planets.  

Take Sun-Earth at 10 pc as a typical case, since the Dep of Earth is around 86 ppm, which is above the precision of S/N = 1 of Kepler. Our results show that the predicted number of Earth-like planets around G stars is mainly limited by both the stellar population and the strict geometric constraint (low TP), rather than the photometric precision.

\subsection{Astrometry}\label{sec:astrodetect}

For the astrometry method, the maximum stellar displacement is adopted as the observable signal, which depends on the distance and the stellar mass. According to equation \ref{astrometry formula},  astrometry prefers finding planets in the HZ around massive stars, as interpreted in Section \ref{sec:astrometryest}. Taking Proxima and the Solar system at 10 pc for instance, the maximum stellar displacement due to Earth of Sun is 0.3$\mu$as, while, the displacement of Proxima at 10 pc is about $0.125\mu$as. 

\begin{figure}
\centering
\includegraphics[width=0.45\textwidth,height=0.45\textwidth]{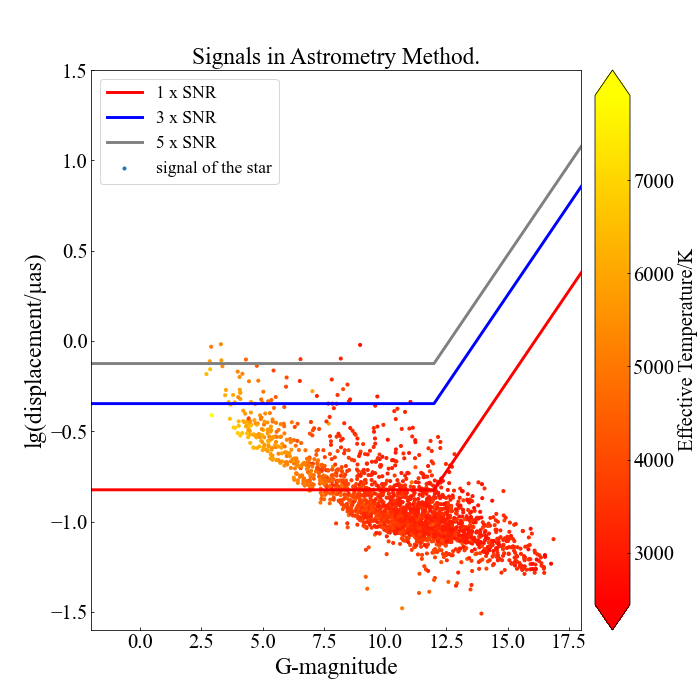}
\caption{The astrometric signals and three different precision models. The red, blue, and grey lines represent the precision of 1.5\%, 0.9\%, and 0.3\% of Gaia's precision. }
\label{fig:Astrometry Expectation}
\end{figure}

Astrometry is already used to detect exoplanets and estimate their parameters \citep{Venner_2021}. Thanks to the high astrometry precision of Gaia, astrometry has found decades of stellar companions in DR3 \citep{2022arXiv220700680A}, as well as some massive planetary companions. For stars with $G_{\rm mag}$ of around 15, the precision of Gaia EDR3 is about 200$\mu$as, while the best precision of 50 $\mu$as is achieved for stars with a $G_{\rm mag}$ between 9.0 and 12.0 \citep{refId0}. Note that Gaia is doing absolute astrometry, while detecting planets only needs relative astrometry, which is usually more accurate than absolute astrometry like Gaia via choosing reference stars. For instance, CHES is expected to detect the displacement at $\geq 1 \upmu$as \citep{Ji_2022}. Moreover, several projects using astrometry to detect nearby planets are about to execute in the future \citep{Janson_2018}, e.g. Theia, which also uses relative astrometry by using high-cadence observations of each target and more than three reference stars \citep{Malbet2022TheiaSC}. In consideration of the parameters of existing instruments, here we adopt a typical precision model (gray line) much higher than Gaia, as shown in Figure \ref{fig:Astrometry Expectation}, with the best precision of 0.75 $\mu$as for stars brighter than $G_{\rm mag}$=12. 


From section \ref{sec:astrometryest}, we have found that more than 98\% of the maximum stellar displacements of stars within 20 pc are under 1 $\mu$as, which means we need to improve the precision to detect more exoplanets. So we assume two additional models with better precision, as shown with red and blue lines in Figure \ref{fig:Astrometry Expectation},  i.e. the best precision is set to 0.45 and 0.15 $\mu$as, respectively. 

According to Figure \ref{fig:Astrometry Expectation}, if we set the detecting criterion as the gray lines, five habitable planets are expected to be detected, including one planet around G stars and two planets around M stars. If we change the criterion to the blue line, i.e. the precision is improved 1.67 times, the expected number increases to 30, including seven planets around G stars and 10 planets around M stars. Furthermore, if we use the most precise astrometry with the criterion of the red line, the expected number becomes 511, including 116 and 174 planets around G and M stars, respectively.

The result shows that, different from the transit and RV methods, astrometry is not preferred for M stars. After we studied the proportion of detectable planets around different types of stars, the result shows that 96.0\% of the planets in the HZ around G-type stars can be detected with the precision of the red line of figure \ref{fig:Astrometry Expectation}. Therefore, the astrometry method needs a precision of $\sim 0.15 \mu$as (red line) to release its great potential to find nearby Earth-like planets in the HZ of solar-like stars. However, detecting planets around M stars requires a much higher astrometric precision of 0.10 $\mu$as.

\subsection{Imaging}\label{sec:imagingdetect}



Imaging has found six nearby exoplanets within 20 pc. The imaging method has a unique advantage over both RV and transit; i.e. it is not sensitive to the inclination of the planets, as the largest angular separation only depends on the distance. Nearby stars close to the Sun are more likely to be distinguished from planets in the HZ. As a typical case, the nearest, Proxima b, has an angular separation of $0.^{''}038$ to the host, while the separation becomes $0.^{''}0025$ if we put the system at 20 pc. 

\begin{figure*}
\centering
\includegraphics[width=0.96\textwidth,height=0.48\textwidth]{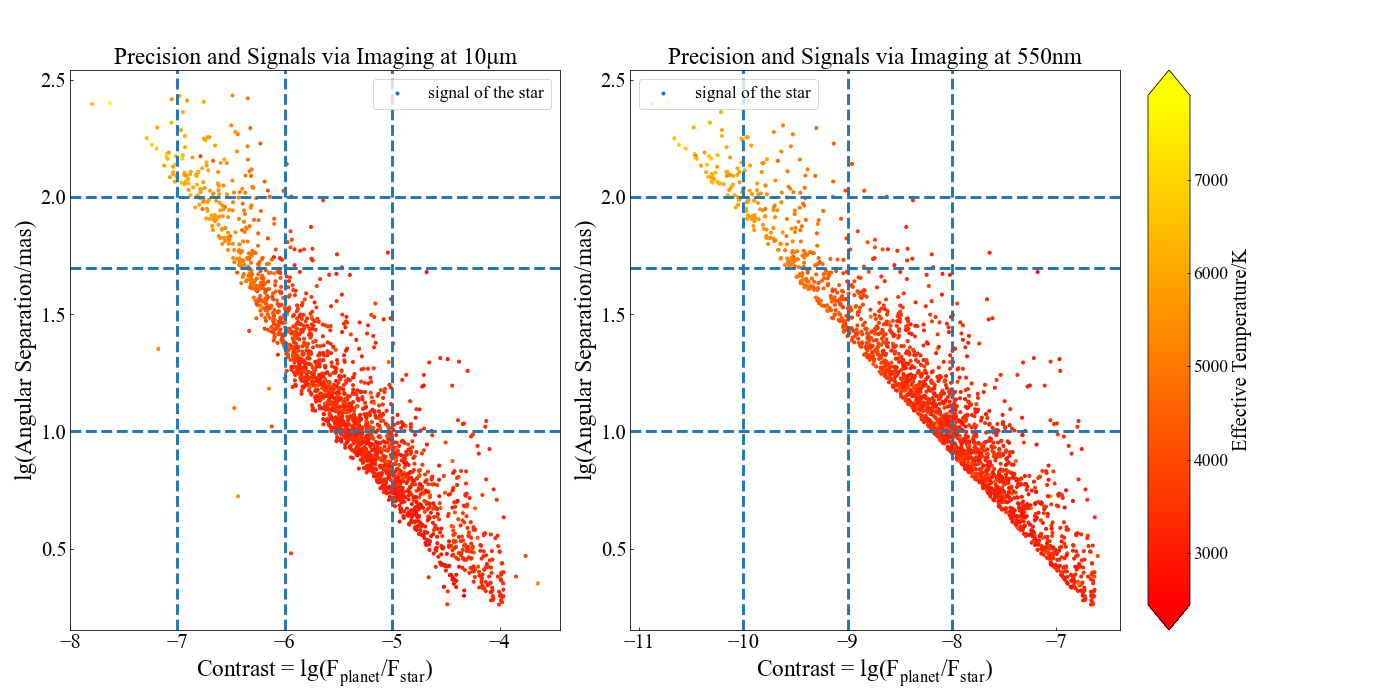}
\caption{Contrast and angular separation between planets and stars in the infrared band (10 $\mu$m; left) and V-band (550 nm i.e. the right). For the infrared band, the contrast is primarily located in the range of $10^{-8}$-$10^{-4}$, so the grid lines represent three different contrasts of $10^{-7}$, $10^{-6}$, and $10^{-5}$ and three different angular resolutions of 100, 50, and 10 mas. For the V band, the contrast is mainly located in the range of $10^{-11}$-$10^{-7}$. The grid lines represent three different contrasts of $10^{-10}$, $10^{-9}$, and $10^{-8}$ and three different angular resolutions of 100, 50, and 10 mas.}
\label{fig:Image detection}
\end{figure*}

According to the Rayleigh criterion, the ability to resolve the planet and star depends on the wavelength and diameter of the aperture. For instance, the NIRCam on JWST has an inner working angle of $0.^{''}14$, using MASKSWB at 2.1 $\mu m$ (see footnote 3), while the MIRI can achieve an inner working angle of $0.^{''}33$ via a four-quadrant phase mask at 10.575 $\mu m$. Another crucial problem for exoplanet imaging is the extreme luminosity ratio. In this subsection, we choose two typical bands: $550nm$ in the optical band and $10\mu m$ in the infrared band. Then we calculate the luminosity ratio and the maximum separations by assuming that the planets are located in the CHZ. The results of the two bands are presented in Figure \ref{fig:Image detection}.


In Figure \ref{fig:Image detection}, we can find out that generally, the angular distance of a planet's orbit generally decreases as the emission ratio between the planet and the star increases. Two extreme regions remain. One region consists of stars with high effective temperatures. The contrast of Earth-like planets to the stars becomes too small to be detectable, as the planets reflect less radiation because of the farther distance from their host stars, and their thermal radiation ratios become small. The other region consists of stars with low effective temperatures. The orbit of the habitable planet becomes too close to distinguish the planets from the host stars, although the contrast becomes larger.
For stars with a temperature of $\leq$ 4000 K, the average contrast at 550nm and angular separation are 1.871$\times10^{-7}$ and 15.97 mas, respectively, while for stars hotter than 6000 K, the average contrast and angular separation are 3.18$\times10^{-10}$ and 151.52 mas.


Compared to the other three methods, the imaging method is relatively complex because it needs to take both contrast and angular resolution into account. Additionally, the imaging method also suffers from weak flux from the planet lot. We assume that the telescopes used to do imaging in the near future will be large enough, e.g., LUVOIR\footnote{https://asd.gsfc.nasa.gov/luvoir/reports/} and HabEx \citep{2020arXiv200106683G} in space and TMT on the ground. Thus, we do not include the flux as a criterion in the following estimation.


We take the host star into consideration and use a criterion of $\geq50$ mas for angular distance. For the optical band, we choose $\geq10^{-10}$ as the criterion of contrast, while for the infrared band, we choose $\geq10^{-7}$ as the criterion. The result is shown in Figure \ref{fig:Image detection}. In the optical band, the predicted number of stars is 159, and 92 of them are G stars. In the infrared band, the predicted number of stars is 191, and 106 of them are around G stars. These detectable planets are around moderate stars, which are neither too hot nor too cool, indicating that imaging has great potential to detect planets around stars like the Sun.


\subsection{Comparison of different methods}

\begin{minipage}[c]{\textwidth}
\centering
\begin{threeparttable}
	\caption{Summary of Precision Models, Detection Criteria, and Expected Earth-like Planets Detected via Four Different Methods. }
	\label{tab:signals of planets, noise model and expectations}
            \footnotesize
            \begin{tabular}{cccccc} 
		\hline
		Method & Observable Signal & Best Precision@mag & Reference Instruments &  Criterion  & Expected Planets \\
		 & (unit) & & (Device) & & (Known, expected, G star)\\
		\hline
		RV & Semi-amplitude(m $s^{-1}$) & 0.2@8 & ESPRESSO & S/N>1 & 4, 39, 1\\
		Transit & Dep(ppm) & 30@7 & TESS & S/N>7  & 4, 5, 0\\
		Image(550$nm$)& Contrast & $>10^{-10}$ & LUVOIR & >$10^{-10}$ and $\theta>50 mas$ & 0,159,92\\
		Image(10$\mu$m) & Contrast & $>10^{-7}$& LIFE\footnote{See \citet{2022AA...664A..21Q}.} & >$10^{-7}$ and $\theta>50 mas$ & 0, 191, 106\\
		Astrometry & $\rm Disp_{\rm Max}(\mu$as) & 0.15@12 & 3xGaia/1000\footnote{The astrometry precision is assumed as 3/1000 of Gaia.} & S/N>3 \footnote{See \citet{2019RAA....19....4Y}} & 0, 30, 8\\
		\hline
	   \end{tabular}
          \begin{tablenotes}
            \item \textbf{Notes.} The three numbers in the last column means the number of known  Earth-like exoplanets, the expectation of Earth-like exoplanets, and the expectation of Earth-like exoplanets around G stars, respectively. The symbol "\@" means that we can reach the best precision on condition that the stars are brighter than a certain magnitude.
          \end{tablenotes}
\end{threeparttable}
\end{minipage}

Based on the detected number of nearby exoplanets in the HZ, the RV method is the most fruitful. Planets around M or K stars, which dominate the nearby stellar populations, can be easily detected compared with those around hotter and larger stars, since the locations of the HZs are closer to the host stars. The most powerful RV instruments, like ESPRESSO, can achieve a precision of 0.1 m/s and are expected to detect dozens of planets around M and K stars. However, if we want to detect exoplanets around G and F stars, the precision of the RV method still needs to be improved. 

The transit method is similar to RV, which is preferred to detect planets around smaller stars. The photometric precision has been pushed to 50 ppm (i.e. Kepler), which is good enough to detect Earth-like planets around M, K, and G stars. However, the geometric constraint limits the total expectations, especially for planets around solar-like stars, which have lower TPs than those around M stars. Additionally, G stars are only $\sim 10\%$ of the nearby stars. Thus, only a few planets around G stars with proper inclinations can be detected, while tens of Earth-like planets around M or K stars are predicted to be detected.  

The astrometry and the imaging methods show great potential to detect Earth-like planets in the HZ around solar-like stars if the precision can be significantly improved. For astrometry, although the signals of most planets in the HZ are below 1$\mu$as, there are six planets around G stars that can be detected with a precision of $1 \mu$as. If the precision can be improved to 0.1 $\mu$as in future missions, Earth-like planets in the HZ around more than 98.0\% of nearby G stars will be detectable. Most Earth-like planets around M stars will be missed with a precision of 0.1 $\mu$as via astrometry but can be complemented via RV and transit. 

Imaging is also sensitive to the angular separation and contrast of the planet and star. The angular separations are typically $\sim$ 10 and 100 mas for planets around G and M stars, respectively. The contrast for Earth-like planets around G stars is about 1-2 orders lower than those around M stars in both the optical and IR bands. Direct imaging in the IR band is more realistic with lower requirements of contrast. Using chronograph (starshade; \citealt{2018NatAs...2..600G}) or nulling interferometry \citep{Quanz_2022}, if we can detect planets with a contrast of $>10^{-7}$, with a separation of $>30 mas$, 93.3\% of the Earth-like planets around G stars will be detectable.



\section{Summary and discussion}\label{sec:sum}

In this paper, we focus on the nearby stars within 20 pc and investigate the potential of detecting Earth-like planets in the HZ. The ability and potential to detect nearby Earth-like planet via different methods can be shown quantitatively based on our results. These results also reveal some advantages and limitations of future space telescope missions aims to detect another Earth.

In section \ref{sec:BPPcollect}, based on the stellar sources in the GCNS, we complete the NSC by collecting and estimating stellar parameters for stars within 20 pc, according to 14 catalogs, including Gaia EDR3, TESS, etc. (Table \ref{tab:catalog list}). We derive the stellar parameters empirically and select 2111 main-sequence stars as the sample in the following simulations (mentioned as the refined NSC), including 1236 M stars, 684 K stars, 150 G stars, 40 F stars, and 1 A star, as shown in Figure \ref{fig:LTcheck}.

In section \ref{sec:HZcalc}, we calculate the extended HZ around nearby stars in Figure \ref{fig:IHZ and OHZ} and conclude that most planets in the HZ with a shorter period than Earth, which benefit from detecting the period signals in a shorter time span. As the central location of the extended HZ we chose is similar to that of the traditional HZ, the extended HZ only enhances the occurrence rate of planets in the HZ, and the average locations will not vary much. The differences occur on the location of the inner and outer boundaries of the HZ. The extended HZ has a larger width and will influence the results of extreme cases; i.e., signals generated by planets located on the inner or outer boundaries are completely different. 

Assuming that every nearby star hosts an Earth-like planet in its HZ, in section \ref{sec:Signal}, we predict the signal distribution of Earth-like planets in the HZ via four methods, i.e., RV, transit, astrometry, and direct imaging, as shown in Figures \ref{fig:RVplot}-\ref{fig:flux-ratio}. The estimation shows that the signal distributions are dominated by the M stars, the dominant population in NSC.

In section \ref{sec:Detect}, we combine the signal distribution and the noise model we adopted to estimate the S/Ns of signals in four methods. For the transit and RV methods, the precision of Kepler/TESS and ESPRESSO is adopted. For direct imaging and astrometry methods, we use the typical precision of future missions, e.g. HabEx and LIFE. An artificial noise model depending on the stellar magnitude is assumed to make comparisons and predictions. According to the noise models, we estimate the total number of detectable Earth-like planets via different methods in our best-case hypothetical scenario, as summarized in Table \ref{tab:signals of planets, noise model and expectations}. The result indicates that the transit and RV methods prefer planets around M stars. The astrometry method prefers G stars and hotter stars directly. Imaging also prefers planets around stars of moderate temperature, i.e. G and K stars. With much higher precision in the future, the RV, astrometry, and direct imaging methods have great potential to detect most Earth-like planets around solar-like stars, while the transit method is limited by the small TP and the small population of nearby stars.

This paper is a preliminary statistical prediction to compare the potential of different methods based on ideal assumptions and the precision of current or proposed future missions. Although we assume that there is one "Earth" in the HZ around every star, this is reasonable compared with the results of \citet{2021AJ....161...36B} and \citet{2021AA...653A.114S}. Exempli gratia, the average number of planets in the HZ around M or G-type stars is around 1.0,  with large uncertainties. The results in this paper will provide an available framework for selecting the targets and designing the precision requirements of the four different methods. Additionally, we do not exclude binaries in the refined NSC because planets around binary stars (P- or S-type) are detectable. Exempli gratia, planets b and c in TOI-1338 are revealed by transit and RV, respectively.
(\citealt{2020AJ....159..253K}, \citealt{2023arXiv230110794S}).

There are still some crucial observational properties that need to be considered. For exmaple, the stellar activity affects the noise to detect Earth-like planets the most in all four methods \citep{Lagrange_2011}. While it is difficult to estimate the influence of stellar activity statistically, time domain photometric and spectral surveys like LSST will help us to understand the activities of nearby stars. Multiplanetary systems also produce extra residual noise to submerge the signal of Earth-like planets in the HZ \citep{2019RAA....19....4Y}. The occurrence rate of planets in the HZ correlated with stellar parameters is also crucial to obtaining a reasonable number. Further prediction for a certain instrument should be simulated case by case using a more realistic noise model. For future space missions, like LUVOIR, HabEx, LIFE, etc., using more developed instruments, we hope the detecting ability can be improved and lead to unique, amazing detections of "Earth twins," i.e., Earth-like planets around nearby solar-like stars.


\section*{Acknowledgements}

This work is supported by the National Natural Science Foundation of China (grant Nos. 11973028, 11933001, 1803012, 12150009), and the National Key R\&D Program of China (2019YFA0706601). We also acknowledge the science research grants from the China Manned Space Project with Nos.CMS-CSST-2021-B12 and CMS-CSST-2021-B09, as well as the Civil Aerospace Technology Research Project (D050105).

We also thank the Guoshoujing Telescope (the Large Sky Area Multi-Object Fiber Spectroscopic Telescope, LAMOST), which is a National Major Scientific Project built by the Chinese Academy of Sciences. Funding for the project has been provided by the National Development and Reform Commission. LAMOST is operated and managed by the National Astronomical Observatories, the Chinese Academy of Sciences.

%

\section*{Data Availability}

The data underlying this article are mainly based on the Gaia Catalog of Nearby Stars (GCNS), public Gaia DR3, and Gaia DR2, as well as other catalogs in the Vizier database (https://vizier.u-strasbg.fr/viz-bin/VizieR-2), as shown in Table 1. The NSC and parameters used in this paper are available in GitHub-NSC (see footnote 2).

\bibliography{PASPsample631}{}
\bibliographystyle{aasjournal}



\end{document}